\documentclass{elsart}
\usepackage{amssymb,epsfig}

\usepackage{graphicx}
\usepackage{bm}

\def\be{\begin{equation}} 
\def\ee{\end{equation}}
\def\bq{\begin{eqnarray}} 
\def\eq{\end{eqnarray}}

\def\fmr{\mathrm {fm}}
\def\MeVr{\mathrm {MeV}}

\begin{document}
\begin{frontmatter}

\title{$\!\!\!$Testing Deconfinement at High Isospin Density}

\author{M. Di Toro$^1$, A. Drago$^2$, T. Gaitanos$^3$, V. Greco$^1$, 
A. Lavagno$^4$}

\address{
$^1$ Universit\`a di Catania and INFN, Lab. Nazionali
del Sud, 95123 Catania, Italy}
\address{
$^2$ Dip. di Fisica, Univ. Ferrara 
and INFN, Sez. Ferrara, 44100 Ferrara, Italy}
\address{
$^3$ Dept. f\"ur Physik, Universit\"at M\"unchen, D-85748 Garching, Germany}
\address{
$^4$ Dip. di Fisica, Politecnico di Torino}
\address{E-mail: ditoro@lns.infn.it}
\date{\today }
\maketitle

\begin{abstract}
We study the transition from hadronic matter to a mixed phase of
quarks and hadrons at high baryon and isospin densities reached 
in heavy ion collisions. We focus our attention 
on the role played by the nucleon symmetry energy at high density.In this
respect the inclusion of a scalar isovector meson, the
$\delta$-coupling, in the Hadron Lagrangian appears rather important.
We study in detail the formation of a drop of quark matter in the
mixed phase, and we discuss the effects on the quark drop nucleation
probability of the finite size and finite time duration of the high
density region.  We find that, if the parameters of quark models are
fixed so that the existence of quark stars is allowed, then the
density at which a mixed phase starts forming drops dramatically in the
range $Z/A \sim $ 0.3--0.4. This opens the possibility to verify the
Witten-Bodmer hypothesis on absolute stability of quark matter using
ground-based experiments in which neutron-rich nuclei are employed.
These experiments can also provide rather stringent constraints on the
Equation of State ($EoS$) to be used for describing the 
pre-Supernova gravitational collapse.
Consistent simulations of neutron rich heavy ion collisions 
are performed in order to show that even at relatively low energies,
in the few $AGeV$ range, the system can enter such unstable mixed phase.
Some precursor observables are suggested, in particular a ``neutron
trapping'' effect.



\end{abstract}
\end{frontmatter}

\section{Introduction}

Hadronic matter is expected to undergo a phase transition 
into a deconfined phase of quarks and gluons at large densities 
and/or high temperatures. On very general grounds,
the transition's critical densities are expected to depend
on the isospin of the system, but no experimental test of this 
dependence has been discussed in the literature.
Up to now, experimental data on the phase transition have been 
extracted from
ultrarelativistic collisions of almost isospin-symmetric nuclei, having
a proton fraction $Z/A\sim$ 0.4--0.5. Moreover,
in those experiments large temperatures are obtained, but the 
maximum density is not much larger than nuclear matter saturation
density $\rho_0$.
Experiments with lower beam energies, in which high 
baryon densities can be reached, have not been
extensively studied with the aim of detecting specific signatures
of the transition. The analysis of observations of neutron stars,
which are composed of $\beta$-stable matter for which
$Z/A\lesssim$ 0.1, can also provide hints on the structure
of extremely asymmetric matter at high density. 
No data on the quark deconfinement transition 
is at the moment available for intermediate values of
$Z/A$. Recently it has been proposed
by several groups to produce unstable neutron-rich beams 
at intermediate energies. 
As we will show,
these new experiments open the possibility to explore in laboratory the isospin
dependence of the critical densities.

The information coming from experiments with relativistic 
heavy ions is that, for symmetric or nearly symmetric nuclear matter, 
the critical density appears to be considerably larger than $\rho_0$.
Concerning non-symmetric matter, general arguments based on Pauli principle
suggest that the critical density decreases with $Z/A$. 
We want to study in particular the range $Z/A\sim$ 0.3--0.4, which can be
partially explored using beams of neutron rich nuclei as $^{238}U$
and more extensively tested
in radioactive nuclear beam facilities.
This region is also 
relevant for supernova explosion.
Here and in the
following we are mainly interested in the transition density 
separating pure hadronic matter from a mixed phase of hadrons
and quarks. The second transition density, separating
the mixed phase from the pure quark matter phase, 
cannot be reached in intermediate energy experiments.

A study of the isospin dependence of the transition densities has been
performed up to now, to our knowledge, only by Mueller
\cite{MuellerNPA618}, although in that work only one set of model parameter
values is explored. The conclusion of \cite{MuellerNPA618} is that,
moving from symmetric nuclei to nuclei having $Z/A\sim 0.3$, the
critical density is reduced by roughly 10$\%$.  In this paper we
explore in a more systematic way the model parameters and we estimate
the possibility of forming a mixed-phase of quarks and hadrons in
experiments at energies of the order of a few $GeV$ per nucleon.
Moreover, as a crucial technical refinement of previous analysis, we
will discuss in detail the formation of a drop of quark matter, taking
into account possible retardation effects associated with a
non-vanishing surface tension at the quark-hadron interface.  Clearly,
since the high density region has a finite size and a finite duration,
an ``effective'' critical (transition) density has to be reached, 
so that the quark drop nucleation rate is not
too small.  We show that quark clusters can indeed be produced
on the expected time scale, at a
density not much larger than the normal spinodal critical density.
The effect of a finite temperature is also taken into account.

Concerning the hadronic phase, we have first used the relativistic
non-linear Walecka-type model of Glendenning-Moszkowski ($GM1$, $GM2$,
$GM3$) \cite{GlendenningPRL18}. This effective field Lagrangian is very
similar to the one of Ref.~\cite{MuellerNPA618}, but with a different
choice of the coupling constants in order to reproduce a softer $EoS$
for symmetric matter at high baryon density. This is in fact more in
agreement with relativistic Heavy-Ion-Collision ($HIC$) data
\cite{DanielNPA673,DanielSCI298} and with correlated
Dirac-Brueckner-Hartree-Fock ($DBHF$)
\cite{BrockmannPRC42,GrossNPA648,HaarPR149,GaitanosEPJA12} results,
see the discussion in
Refs.~\cite{GrecoPLB562,GaitanosNPA732,GaitanosPLB595}.   
The isovector part is treated analogously to the isoscalar part,
by introducing a coupling to a vector charged meson.
To further explore the sensitivity of our
results on the hadronic $EoS$, a large part of our work is devoted to
investigating the possibility of enhancing the symmetry repulsion at
high baryon density by introducing a coupling to a charged scalar
$\delta$-meson. As remarked in Ref.\cite{LiuboPRC65}, this is fully in
agreement with the spirit of effective field theories, and of
course with the phenomenology of the free nucleon-nucleon interaction.

For the quark phase we have considered the $MIT$ bag model
\cite{MitbagPRD9} at first order in the strong coupling constant
$\alpha_s$ \cite{FarhiPRD30,MuellerNPA618}.  In order to show the
soundness of the discussed effects, in some cases we have repeated the
calculations also using the Color Dielectric Model ($CDM$)
\cite{Birse1990,Pirner1992,Drago1995} for the quark phase.  In the
latter, quarks develop a density dependent constituent mass through
their interaction with a scalar field representing a multi-gluon
state.  In particular we will be interested in those parameter sets
which would allow the existence of quark stars
\cite{AlcockAJ310,HaenselAA160,DragoPLB511}, i.e. parameters sets for
which the so-called Witten-Bodmer hypothesis is satisfied
\cite{WittenPRD30,BodmerPRD4}.  According to that hypothesis, a state
made of an approximately equal number of up, down and strange quarks
can have an energy per baryon number $E/A$ smaller than that of iron
($E_{Fe}\approx 930~MeV$).  To satisfy the Witten-Bodmer hypothesis,
strong constraints on quark model parameters have to be imposed. For
instance, using the basic version of the $MIT$ bag model, the so-called
pressure-of-the-vacuum parameter $B$ must have a very small value,
$B^{1/4}\sim$ 140--150 $MeV$ \cite{FarhiPRD30,DragoPLB511}.  Taking into
account the possibility of forming a diquark condensate, quark stars
can exist also for larger values of the pressure of the vacuum, up to
$B^{1/4}\sim$ 160--180 $MeV$ \cite{AlfordPRD67,DragoPRD69}, depending on
the type and size of the superconducting gap.  These are the largest
values of $B$ that we will discuss in our analysis, since one of the
aim of our paper it to show that if quark stars are indeed possible,
it is then very likely to find signals of the formation of a mixed
quark-hadron phase in intermediate-energy heavy-ion experiments.
Assuming Witten-Bodmer hypothesis to be true, ordinary nuclear matter
would be metastable. In order not to contradict the obvious stability
of normal nuclei, quark matter made of only two flavors must not be
more stable than iron. Slightly more strict boundaries on parameters'
value can be imposed by requiring not only iron, but also neutron rich
nuclei like e.g. lead, to be stable.  If the Witten-Bodmer hypothesis
is satisfied, self-bound stars entirely composed of quark matter can
exist \cite{AlcockAJ310,HaenselAA160,DragoPLB511}.  Recently several
analysis of observational data have emphasized the possible existence
of compact stars having very small radii, of the order of 9 kilometers
or less \cite{LiPRL83,DeyPLB438,PonsAJ564,DrakeAJ572}.  The most
widely discussed possibility to explain the observed mass-radius
relation is based on the existence of quark stars.  It is therefore
particularly interesting to envisage laboratory experiments testing
the possible signatures of model parameters values that would allow
the existence of these extremely compact stellar objects. What we are
proposing in this paper is to use beams of neutron-rich nuclei to this
purpose.

It is rather unlikely, at least in the near future, that neutron rich
nuclei obtainable in radioactive beam facilities can be accelerated to
very large energies, much larger than a few $GeV$ per nucleon. On the
other hand, these energies are sufficient to our purposes.  The
scenario we would like to explore corresponds to the situation
realized in experiments at moderate energy, in which the temperature
of the system is at maximum of the order of a few ten $MeV$.  In this
situation,
only a tiny amount of strangeness can be produced and therefore in
this paper we only study the deconfinement transition from nucleonic
matter into up and down quark matter.

After having chosen a model for the hadronic and for the quark $EoS$,
the deconfinement phase transition is then described by imposing Gibbs
equilibrium conditions \cite{Landaustat,GlendenningPRD46}.

In order to check if the mixed phase region can be reached in realistic
heavy ion collisions at relativistic energies we have performed
``ab initio'' reaction simulations using relativistic transport
equations. With the same effective lagrangians discussed above
we have analyzed collisions of neutron rich
nuclei, as $^{132}Sn+^{132}Sn$ and the less exotic  $^{238}U+^{238}U$,
at $1~AGeV$ beam energies.
Since at this low energy the interacting system enters
only marginally the mixed phase, we have devoted a whole 
section
to discuss the nucleation mechanism for quark cluster formation, typical of
the metastable regions. We show that quark clusters can be produced
with this mechanism even on a short time-scale, of the order of
$ \simeq 10fm/c$, expected for the lifetime
of a transient state of very exotic interaction matter formed
during the reaction dynamics.

Due to the different relevance of the symmetry repulsion in the hadronic
and quark phases we observe a clear ``neutron distillation'' to the quark 
clusters ({\it neutron trapping} effect), in particular just above the 
transition density. This suggests
a series of possible observables rather sensitive to the deconfinement
transition at high baryon density.

\section{Equations of State}
\label{EOS}
\subsection*{Hadronic Matter}

A Relativistic Mean Field ($RMF$) approach to nuclear matter with the
coupling to an isovector scalar field, a virtual $a_{0}(980)$
$\delta$-meson, has been studied for asymmetric nuclear matter at
low densities, including its linear response
\cite{KubisPLB399,LiuboPRC65,GrecoPRC67}, and for heavy ion collisions
at intermediate energies, where larger density and momentum regions
can be probed, \cite{GrecoPLB562,GaitanosNPA732,GaitanosPLB595}.  In
this work we extend the analysis of the contribution of the
$\delta$-field in dense asymmetric matter to the transition to a
deconfined phase.

A Lagrangian density of the interacting many-particle system
consisting of nucleons, isoscalar (scalar $\sigma$, vector $\omega$),
and isovector (scalar $\delta$, vector $\rho$) mesons is the starting
point of our $RMF$ approach. We will call this the
$Non-Linear(\rho,\delta)$ model, $NL\rho$ and $NL\rho\delta$ :

\begin{eqnarray}\label{eq:1}
{\mathcal L } &=& \bar{\psi}[i\gamma_{\mu}\partial^{\mu}-(M-
g_{\sigma}\phi -g_{\delta}\vec{\tau}\cdot\vec{\delta})
-g{_\omega}\gamma_\mu\omega^{\mu}-g_\rho\gamma^{\mu}\vec\tau\cdot
\vec{b}_{\mu}]\psi \nonumber \\&&
+\frac{1}{2}(\partial_{\mu}\phi\partial^{\mu}\phi-m_{\sigma}^2\phi^2)
-U(\phi)+\frac{1}{2}m^2_{\omega}\omega_{\mu} \omega^{\mu}
+\frac{1}{2}m^2_{\rho}\vec{b}_{\mu}\cdot\vec{b}^{\mu} \nonumber
\\&&
+\frac{1}{2}(\partial_{\mu}\vec{\delta}\cdot\partial^{\mu}\vec{\delta}
-m_{\delta}^2\vec{\delta^2}) -\frac{1}{4}F_{\mu\nu}F^{\mu\nu}
-\frac{1}{4}\vec{G}_{\mu\nu}\vec{G}^{\mu\nu},
\end{eqnarray}
where ($\phi$, $\omega_{\mu}$) are the isoscalar (scalar, vector) meson
fields, while the ($\vec{\delta}$, $\vec{b}_{\mu}$) are the corresponding 
isovector ones.
$F_{\mu\nu}\equiv\partial_{\mu}\omega_{\nu}-\partial_{\nu}\omega_{\mu}$,
$\vec{G}_{\mu\nu}\equiv\partial_{\mu}\vec{b}_{\nu}-\partial_{\nu}\vec{b}_{\mu}$,
and the $U(\phi)$ is a nonlinear potential of $\sigma$ meson :
$U(\phi)=\frac{1}{3}a\phi^{3}+\frac{1}{4}b\phi^{4}$.
We remind that the  Glendenning-Moszkowski 
($GM1$, $GM2$, $GM3$) \cite{GlendenningPRL18} Lagrangians have exactly 
the same form,
but for the $\delta$-field contribution.

The field equations in $RMF$ approximation are

\begin{eqnarray}\label{eq:2}
&&(i\gamma_{\mu}\partial^{\mu}-(M- g_{\sigma}\phi
-g_\delta{\tau_3}\delta_3)-g_\omega\gamma^{0}{\omega_0}
-g_\rho\gamma^{0}{\tau_3}{b_0})\psi=0,\nonumber \\
&&m_{\sigma}^2\phi+ a{{\phi}^2}+ b{{\phi}^3}=g_\sigma<\bar\psi\psi>=g_\sigma{\rho}_s, \nonumber \\
&&m^2_{\omega}\omega_{0}=g_\omega<\bar\psi{\gamma^0}\psi>=g_\omega\rho,
\nonumber \\
&&m^2_{\rho}b_{0}=g_\rho<\bar\psi{\gamma^0}\tau_3\psi>=g_\rho\rho_3,
\nonumber \\
&&m^2_{\delta}\delta_3=g_{\delta}<\bar\psi\tau_3\psi>=g_{\delta}\rho_{s3},
\end{eqnarray}
where $\rho_3=\rho_p-\rho_n$ and
$\rho_{s3}=\rho_{sp}-\rho_{sn}$, $\rho$ and $\rho_s$ are
the baryon and the scalar densities, respectively.


Neglecting the derivatives of mesons fields, the energy-momentum tensor 
is given by

\begin{eqnarray}\label{eq:3}
\!\!\!\!T_{\mu\nu} = i\bar{\psi}\gamma_{\mu}\partial_{\nu}\psi+[\frac{1}{2}
 m_{\sigma}^2\phi^2+U(\phi)+\frac{1}{2}m_{\delta}^2\vec{\delta^2}
-\frac{1}{2}m^2_{\omega}\omega_{\lambda} \omega^{\lambda}
-\frac{1}{2}m^2_{\rho}\vec{b_{\lambda}}\vec{b^{\lambda}}]g_{\mu\nu}.
\end{eqnarray}


The $EoS$ for nuclear matter with the isovector scalar field at finite
temperature in $RMF$ is given by the
energy density

\noindent
\begin{eqnarray}\label{eq:4}
\epsilon &=& 2 \sum_{i=n,p}\int \frac{{\rm
d}^3k}{(2\pi)^3}E_{i}^*(k) (n_{i}(k)+\bar{n}_{i}(k))
+\frac{1}{2}m_\sigma^{2}\phi^2 \nonumber \\&&
+ U(\phi)
+\frac{1}{2}m_\omega^{2}\omega_0^2
+\frac{1}{2}m_\rho^{2}b_0^2
+\frac{1}{2}m_\delta^{2}\delta_{3}^{2}~,
\end{eqnarray}
and pressure 

\noindent
\begin{eqnarray}\label{eq:5}
 P &=& \frac{2}{3}\sum_{i=n,p}\int \frac{{\rm d}^3k}{(2\pi)^3}
\frac{k^2}{E_{i}^*(k)} (n_{i}(k)+\bar{n}_{i}(k))
-\frac{1}{2}m_\sigma^2\phi^2 \nonumber \\&&
-U(\phi)+\frac{1}{2}m_\omega^2\omega_0^2 +\frac{1}{2}m_{\rho}^2{b_0^2}
-\frac{1}{2}m_\delta^{2}\delta_{3}^{2}~,
\end{eqnarray}
where ${E_i}^*=\sqrt{k^2+{{m_i}^*}^2}$. The nucleon effective masses
are defined as

\noindent
\begin{equation}\label{eq.6}
{m_i}^*=M-g_\sigma\phi\mp g_\delta\delta_3~~~ (-~proton, +~neutron).
\end{equation}

The $n_i(k)$ and $\bar{n}_{i}(k)$  are the fermion and antifermion
distribution functions for protons ($i=p$) and neutrons ($i=n$):

\noindent
\begin{eqnarray}\label{eq.7}
n_i(k)=\frac{1}{1+\exp\{({E_i}^*(k)-{\mu_i^*})/T \} }\,,
\end{eqnarray}

\noindent
and

\begin{eqnarray}\label{eq.8}
\bar{n}_{i}(k)=\frac{1}{1+\exp\{({E_i}^*(k)+{\mu_i^*})/T \} }.
\end{eqnarray}

\noindent
where the effective chemical potential ${\mu_{i}}^*$
is determined by the nucleon density

\noindent
\begin{eqnarray}\label{eq:9}
\rho_i=2\int\frac{{\rm d}^3k}{(2\pi)^3}(n_{i}(k)-\bar{n}_{i}(k))\,,
\end{eqnarray}

\noindent
and the $\mu_i^*$ is related to the chemical potential $\mu_i$ in terms
of the vector meson mean fields by the equation

\noindent
\begin{eqnarray}\label{eq.10}
\mu_i^* ={\mu_i} - g_\omega\omega_0\mp g_{\rho}b_0~~~(-~proton, +~neutron),
\end{eqnarray}

\noindent
where $\mu_i$ are the thermodynamical chemical potentials
$\mu_i=\partial\epsilon/\partial\rho_i$. At zero temperature they
reduce to the Fermi energies $E_{Fi} \equiv \sqrt{k_{Fi}^2+{m_i^*}^2}$.

The proton and neutron chemical potentials can be written in terms of
the baryon and isospin chemical potentials by the equations
\begin{equation}
\mu_p=\mu_B+\mu_3\, , \ \ \ \ \ \ \ \mu_n=\mu_B-\mu_3\, .
\end{equation}

\noindent
The scalar density $\rho_{s}$ is given by

\noindent
\begin{eqnarray}\label{eq:11}
\rho_s=2\sum_{i=n,p}\int\frac{{\rm d}^3k}{(2\pi)^3}\frac{m_{i}^{*}}{E_{i}^*}
(n_{i}(k)+ \bar{n}_{i}(k))~.
\end{eqnarray}

\noindent
where the Fermi momentum $k_{F_i}$ of the nucleon is related to its
density, $k_{F_i}=(3\pi^2 \rho_i)^{1/3}$.

In the presence of a coupling to an isovector-scalar $\delta$-meson field,
 the expression for the symmetry energy at $T=0$ has a simple transparent
form, see \cite{LiuboPRC65,GrecoPRC67}:

\begin{equation}\label{eq:12}
E_{sym}(\rho)=\frac{1}{6} \frac{k_{F}^{2}}{E_{F}}+
\frac{1}{2}[f_{\rho}-f_{\delta}(\frac{m^*}{E_{F}^*})^{2}]\rho~,
\end{equation}

\noindent
where $m^{*}=M-g_{\sigma}\phi$ and
${E_F}^*=\sqrt{k_{F}^2+{m^*}^2}$. We clearly see the mechanism which
is behind the apparent paradox of an attracting contribution of the
isovector scalar field leading to a larger repulsion of the symmetry
term. In fact, at normal density a larger $\rho-meson$ coupling is needed
in order to reproduce the correct symmetry energy coefficient of the
Bethe-Weisz\"acker mass formula.  When the baryon density 
increases, the $\delta$ contribution is quenched by the
$(m^*/E_{F}^*)^{2}$ factor and we are left with a stiffer
symmetry term.  The isovector coupling constants, 
both in the $NL\rho$ and in the $NL\rho\delta$ cases, 
are fixed from the symmetry energy at saturation
and from Dirac-Brueckner estimations, see the detailed discussions in
Refs. \cite{LiuboPRC65,GrecoPRC67}.

It is interesting to compare the predictions on the transition to a
deconfined phase of the two effective Lagrangians $GM3$ and
$NL\rho\delta$.  The isoscalar part is very similar 
in the two models
and at high densities it approaches
Dirac-Brueckner predictions, as already noted.  The isovector part is
quite different, because in $GM3$ we only have the coupling to the vector
$\rho$-field, while in $NL\rho\delta$ we also have the contribution of the
$\delta$-field, which leads to a stiffer symmetry term and to a
neutron/proton effective mass splitting.  We remind that recently the
latter interaction has been used with success to describe reaction
observables in $RMF$-transport simulations of relativistic heavy ion
collisions, where high densities and momenta are reached
\cite{GrecoPLB562,GaitanosNPA732,GaitanosPLB595}.

The coupling constants, $f_{i}\equiv g_{i}^{2}/m_{i}^{2}$, $i=\sigma,
\omega, \rho, \delta$, and the two parameters of the $\sigma$
self-interacting terms: $A_\sigma\equiv a/g_{\sigma}^{3}$ and 
$B_\sigma\equiv b/g_{\sigma}^{4}$ for the two hadron effective interactions are
reported in Tab.~1.  The corresponding properties of nuclear matter
are listed in Tab.~2.

\begin{table}[htb]
\vspace{0.5cm}
\begin{center}
\centerline{{\bf Table 1.} Parameter sets.}

\begin{tabular}{c|c|c|c|c|c} 
Parameter  
          &$NL\rho$       &$NL\rho\delta$ & $GM1$ & $GM2$   & $GM3$   \\ \hline
$f_\sigma~(\fmr^2)$ &10.329   &10.329 & 11.79 & 9.148  &9.923  \\ \hline
$f_\omega~(\fmr^2)$ &5.423   &5.423   & 7.149 & 4.82  &4.82  \\ \hline
$f_\rho~(\fmr^2)$  &0.95   &3.150     & 1.103 & 1.198 &1.198    \\ \hline
$f_\delta~(\fmr^2)$&0.00      &2.500  & 0.00 & 0.00   &0.00          \\ \hline
$A_\sigma~(\fmr^{-1})$     &0.033  &0.033    & 0.014 & 0.016  &0.041         \\ \hline
$B_\sigma$              &-0.0048  &-0.0048 & -0.001 & 0.013 &-0.0024      \\ \hline
\end{tabular}

\end{center}
\end{table}

\begin{table}[htb]
\vspace{0.75cm}
\begin{center}
\centerline{{\bf Table 2.} Saturation properties of nuclear matter.}

\begin{tabular}{ c|c|c|c|c } 
$    $    &$NL\rho,NL\rho\delta$  & $GM1$ & $GM2$   &   $GM3$  \\ \hline
$\rho_{0}~(\fmr^{-3})$ &0.160  & 0.153 & 0.153 &0.153  \\ \hline
$E/A ~(\MeVr)$         &-16.0 & -16.3 & -16.3 &-16.3 \\ \hline
$K~(\MeVr)$            &240.0 & 300.0 & 300.0 &240.0  \\ \hline
$E_{sym}~(\MeVr)$      &31.3  & 32.5 & 32.5 &32.5 \\ \hline
$M^{*}/M $           &0.75  & 0.70 &  0.78 &0.78  \\ \hline
\end{tabular}

\end{center}
\end{table}

\subsection*{Quark Matter}

In our calculations we will use both a ``minimal'' version of the
$MIT$ Bag model \cite{MitbagPRD9}, in which the interaction inside the
bag is neglected, and also a model
taking into account corrections at first order in
the strong coupling constant $\alpha_s$ \cite{FarhiPRD30,MuellerNPA618}.
We will also display results obtained using the CDM
\cite{Birse1990,Pirner1992,Drago1995}.  We will limit our study to the
two-flavor case $(q=u,d)$. As already remarked in the introduction,
this appears well justified for the application to heavy ion
collisions at relativistic (but not ultra-relativistic) energies. The
fraction of strangeness produced at these energies is very small
\cite{FeriniNPA762}.  In our analysis we have not taken into account
the possibility of forming a diquark condensate. Clearly, the
superconducting gaps accessible in the scenario we are discussing are
the ones pairing up and down quarks only. Moreover, the gap can be
suppressed in the reaction case for several reasons: finite size of the system
\cite{AmorePRD65}, different values of the up and down chemical
potential (particularly so for the strongly asymmetric matter we are
discussing in our paper) and finally for the relatively high temperature
always reached in the high density stage of a reaction (see Sects.IV and V).

The energy density, the pressure and the number density for the quark
$q$ read:

\noindent
\begin{eqnarray}\label{edensbag}
\epsilon= 3 \times 2 \sum_{q=u,d}\int \frac{{\rm d}^3k}
{(2\pi)^3}\sqrt{k^{2}+m_{q}^{2}}(n_{q}+\bar{n}_{q}) +B~,
\end{eqnarray}

\begin{eqnarray}\label{pressbag}
 P =\frac{ 3\times 2}{3}\sum_{q=u,d}\int \frac{{\rm d}^3k}{(2\pi)^3}
\frac{k^2}{\sqrt{k^{2}+m_{q}^{2}}} (n_{q}+\bar{n}_{q}) -B~,
\end{eqnarray}

\begin{eqnarray}\label{qdens}
\rho_{i} = {3\times 2} \int \frac{{\rm d}^3k}{(2\pi)^3}
 (n_{i}-\bar{n}_{i})~,~~~~~i = u, d~;
\end{eqnarray}

\noindent
where $B$ denotes the bag pressure, $m_{q}$ the quark masses, and
$n_q, \bar n_q$ indicate the Fermi distribution functions for quarks and
antiquarks respectively:

\noindent
\begin{eqnarray}\label{nq}
n_q=\frac{1}{1+\exp\{({E_q}-{\mu_q})/T \} }\,,
\end{eqnarray}

\noindent
and

\noindent
\begin{eqnarray}\label{nbarq}
\bar{n}_{q}=\frac{1}{1+\exp\{({E_q}+{\mu_q})/T \} }.
\end{eqnarray}

\noindent
Here $E_{q}=\sqrt{k^{2}+m_{q}^{2}}$ and $\mu_{q}$ are the chemical potentials
for quarks and antiquarks of type $q$. The latter are related to the baryon and 
isospin chemical potential 
\begin{equation}\label{chemihq}
\mu_u = \frac{1}{3}\mu_B + \mu_3\, ,~~~~~\mu_d = \frac{1}{3}\mu_B - \mu_3~.
\end{equation}

The quark densities are related to the baryon and isospin densities
by the following equations
\begin{equation}
\rho_B=\frac{\rho_u+\rho_d}{3}\, , \ \ \ \ \ \ \ \rho_3=\rho_u-\rho_d\, .
\end{equation}

We have considered massless $u,d$ quarks and a range of bag
constants, between $B=(140\; MeV)^4$ and $B=(170\; MeV)^4$. 
These are smaller values 
than the one used in Ref.\cite{MuellerNPA618},
$B=(190\; MeV)^4$. On the other hand this parameter range covers almost
completely the one which can give origin to quark stars, even taking
into account the formation of a diquark condensate.  
As already mentioned, we also present results obtained
taking into account corrections at first order in $\alpha_s$.
Explicit formulae for the contribution of the gluon exchange
to energy and pressure can be found e.g. in Ref.~\cite{glendenningbook}.

No density
dependence for the bag pressure has been introduced, but in the $CDM$,
that we have also explored, something similar to a density dependent
$B$ exists, namely the contribution of the scalar field mimicking a
multi-gluon state. 
The Lagrangian of the CDM reads:
\begin{eqnarray}
     {L} &=& i\bar \psi \gamma^{\mu}\partial_{\mu} \psi 
 +{1\over 2}{\left(\partial_\mu\sigma \right)}^2
     +{1\over 2}{\left(\partial_\mu\vec\pi\right)}^2
     -U\left(\sigma ,\vec\pi\right)   \nonumber\\
     &+&\!\!\!\sum_{f=u,d} {g_f\over f_\pi \kappa} \, \bar \psi_f\left(\sigma
     +i\gamma_5\vec\tau\cdot\vec\pi\right) \psi_f   \nonumber 
      +{1\over 2}{\left(\partial_\mu\kappa\right)}^2
      -{1\over 2}{M}^2\kappa^2 \, , 
\end{eqnarray}
where $U(\sigma ,\vec\pi)$ is the ``mexican-hat'' potential, as in
Ref.\ \cite{neuber}. 

The coupling constants are given by $g_{u,d}=g (f_{\pi}\pm \xi_3)$, 
where $f_{\pi}=93~MeV$ is the pion 
decay constant and $\xi_3=f_{K^\pm}-f_{K^0}=-0.75~ MeV$. 
These coupling constants depend on a single parameter $g$.
Confinement is obtained {\it via} the effective quark masses
$m_{u,d}=-g_{u,d} \bar\sigma/(\bar\kappa f_\pi)$
which diverge outside the nucleon.
Working at mean-field level, the only free parameter is actually the
product $G=\sqrt{g {M}}$. 
In our calculations we have assumed $M$=1.7 $GeV$ and we have explored
various values for $g$.
In this model the last term of the
lagrangian plays a role similar to the vacuum pressure constant $B$
of the MIT bag model. At variance with the latter, in the CDM
the vacuum pressure is density dependent, as anticipated.

\subsection*{Mixed phase}

The structure of the mixed phase is obtained by
imposing the Gibbs conditions \cite{Landaustat,GlendenningPRD46} for
chemical potentials and pressure and by requiring
the conservation of the total baryon and isospin densities
\begin{eqnarray}\label{gibbs}
&&\mu_B^{(H)} = \mu_B^{(Q)}\, , \nonumber \\
&&\mu_3^{(H)} = \mu_3^{(Q)} \, , \nonumber \\ 
&& P^{(H)}(T,\mu_{B,3}^{(H)}) = P^{(Q)} (T,\mu_{B,3}^{(Q)})\, ,\nonumber \\
&&\rho_B=(1-\chi)\rho_B^H+\chi\rho_B^Q \, ,\nonumber \\
&&\rho_3=(1-\chi)\rho_3^H+\chi\rho_3^Q\, , 
\end{eqnarray}
where $\chi$ is the fraction of quark matter in the mixed phase.
In this way we get the $binodal$ surface which gives the phase coexistence region
in the $(T,\rho_B,\rho_3)$ space
\cite{GlendenningPRD46,MuellerNPA618}. For a fixed value of the
conserved charge $\rho_3$, related to the proton fraction $Z/A \equiv
(1+\rho_3/\rho_B)/2$, we will study the boundaries of the mixed phase
region in the $(T,\rho_B)$ plane. We are particularly interested in
the lower baryon density border, i.e. the critical/transition density
$\rho_{cr}$, in order to check the possibility of reaching such
$(T,\rho_{cr},\rho_3)$ conditions in a transient state during a $HIC$
at relativistic energies.

In the hadronic phase, if a quadratic form is assumed for the symmetry
energy the latter is related to the charge chemical potential by the
equation:
\begin{equation}\label{musym}
\mu_3 = 2 E_{sym}(\rho_B) \frac{\rho_3}{\rho_B}\, .
\end{equation} 
We expect therefore that our results on the critical density will be
rather sensitive to the isovector channel in the hadronic $EoS$ at high
densities.

\section{Results at Zero Temperature}
\label{zerotemp}

\begin{figure}
\begin{center}
\includegraphics[scale=0.6]{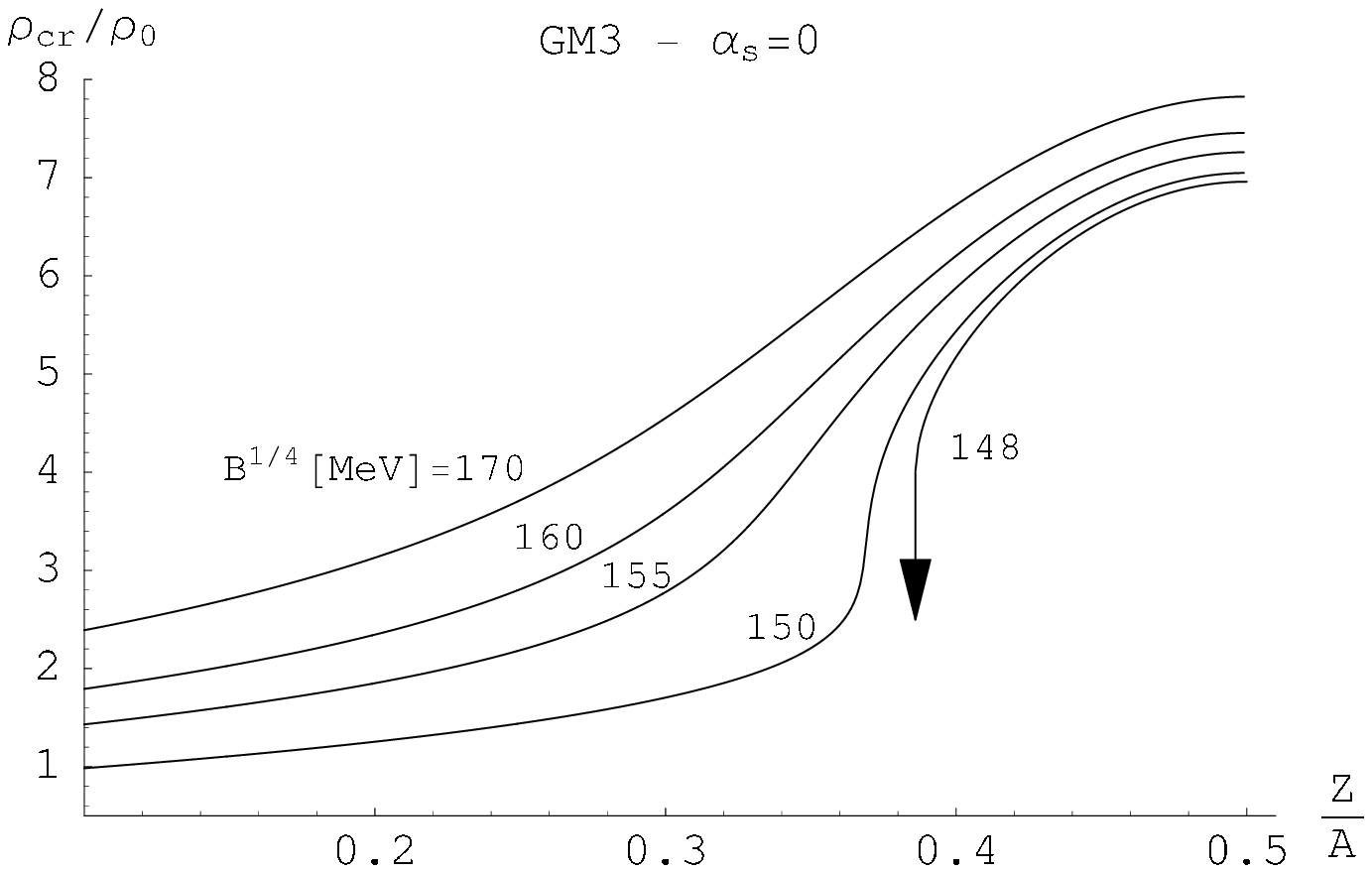}\\
\includegraphics[scale=0.6]{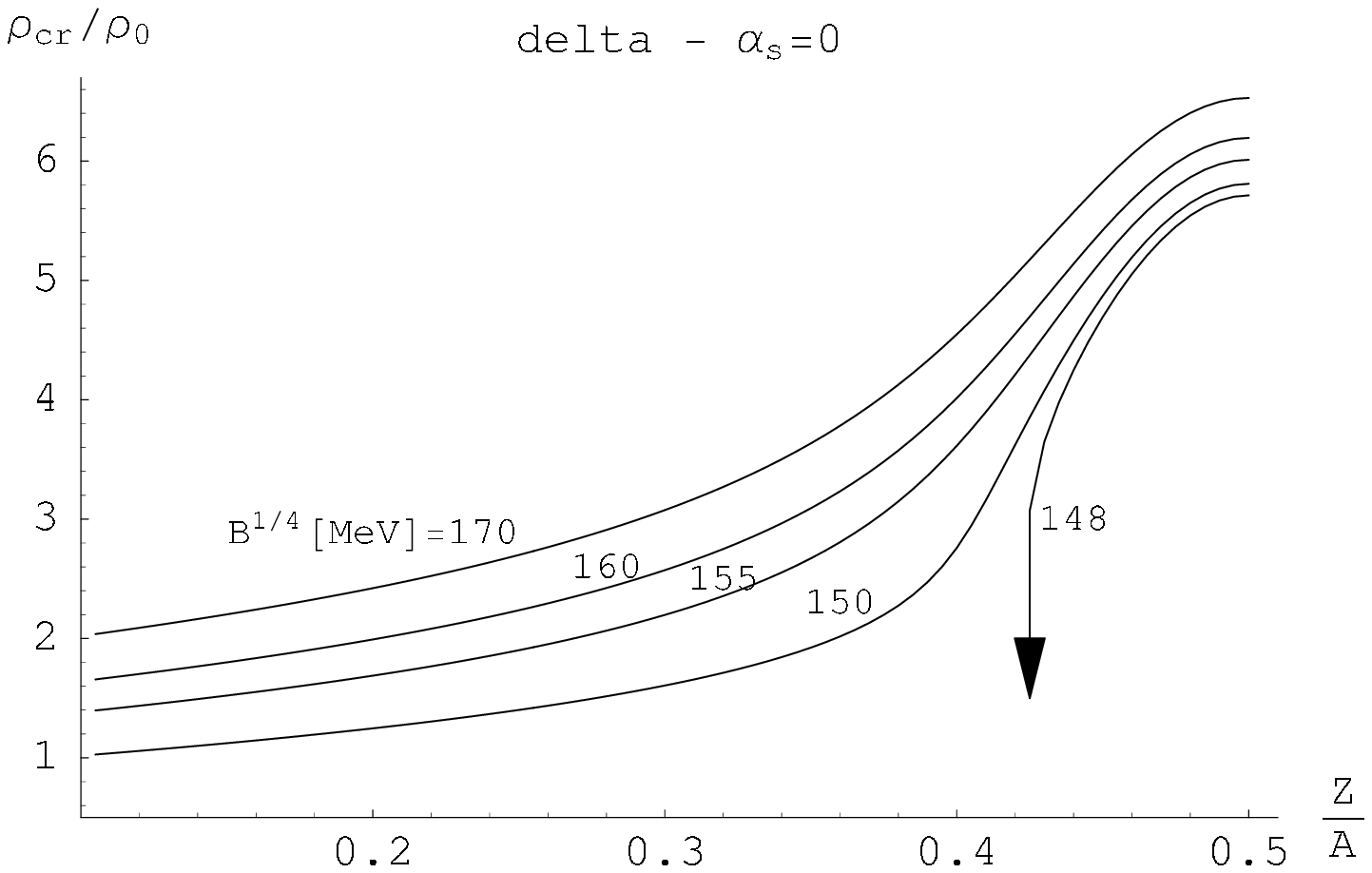}
\end{center}
\caption{\label{fig1}
Transition densities separating hadronic matter from mixed
quark-hadron phase at zero temperature. In the upper panel the $GM3$
parametrization has been used for the hadronic $EoS$, in the lower
panel, $NL\rho\delta$ parametrization, see Table 1.  The $MIT$ bag
model without gluon exchange has been used for the quark $EoS$. The
arrows indicate that the transition density drops to very small values
and the parameter $B^{1/4}$ cannot be further reduced.
}
\end{figure}

\begin{figure}
\begin{center}
\includegraphics[scale=0.7]{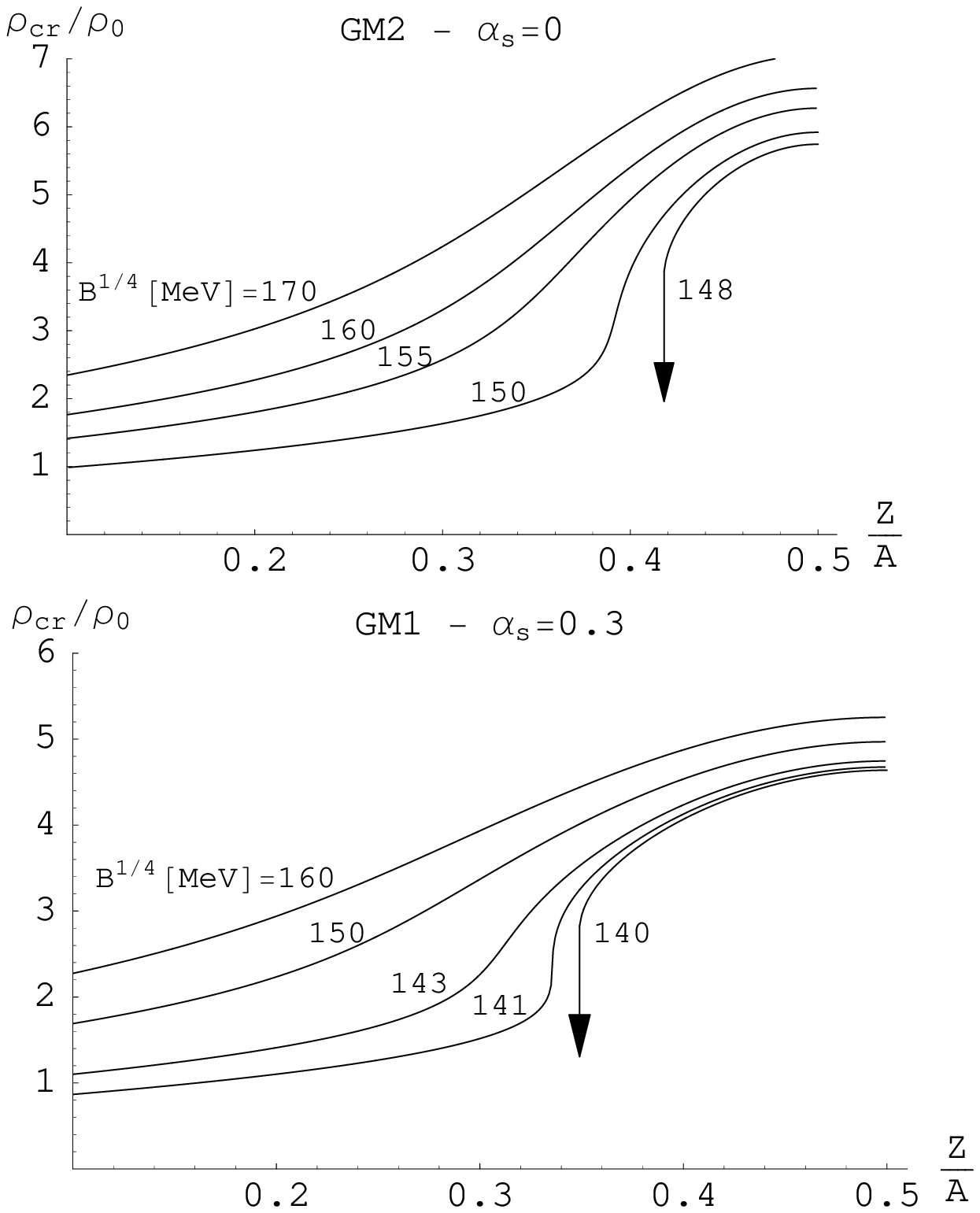}
\end{center}
\caption{\label{gluon}
Similar to Fig.\ref{fig1}. In the upper panel the $GM2$ parametrization
\cite{GlendenningPRL18}
has been used for the hadronic $EoS$ and the $MIT$ bag model without
gluon exchange has been used for the quark $EoS$. In the lower panel,
$GM1$ parametrization \cite{GlendenningPRL18}
for the hadronic $EoS$ and $MIT$ bag model with perturbative 
exchange of gluons with $\alpha_s=0.3$. 
}
\end{figure}

The most plausible way of testing in terrestrial
laboratories the possible formation of a mixed
phase of hadrons and quarks at high baryon and isospin
densities, is via heavy-ion collisions at
relativistic energies, as it will be discussed in Secs. IV--VII.
It is anyway interesting to investigate the possibility of detecting
modifications in the structure of nuclei even with experiments testing
densities of the order of $\rho_0$. This is what is discussed in the
present section.

\begin{figure}
\begin{center}
\includegraphics[scale=0.7]{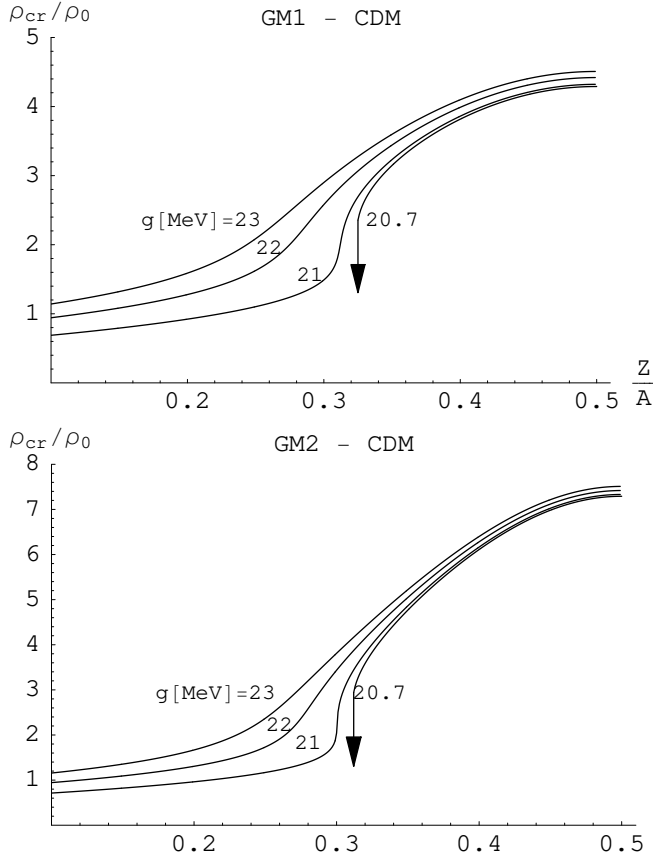}
\end{center}
\caption{\label{cdm}Similar to Fig.1. The
quark $EoS$ has been computed using the Color Dielectric Model
\cite{Drago1995}. The parameter $g$ regulates the coupling
between quarks and a scalar multi-gluon field, see text.}
\end{figure}

In Figs.~\ref{fig1}, \ref{gluon}, \ref{cdm} we report the crossing
density $\rho_{cr}$ separating nuclear matter from the quark-nucleon
mixed phase, as a function of the proton fraction $Z/A$ for various
choices of the Hadronic/Quark $EoS$.  Fig.~\ref{fig1}
shows the $GM3$ (top panel) vs. $NL\rho\delta$ (bottom panel) coupled
to the same no-gluon $MIT$ bag model.  We can see the effect of the
$\delta$-coupling towards an $earlier$ crossing due to the larger
symmetry repulsion at high baryon densities. Fig.~\ref{gluon}
has been obtained using $GM2$ ($GM1$) parametrization for the hadronic
phase and the $MIT$ bag model without gluons (with gluons and
$\alpha_s=0.3$) in the upper and lower window, respectively.  Finally
in Fig. \ref{cdm} the same analysis is performed using the $CDM$ for
the quark $EoS$. For values of the parameter $g$ slightly smaller than
the one indicated in the figure the critical density drops to a very
small value and the situation depicted in Fig.\ref{fig1} using arrows
is obtained.

The most striking feature of all results is the sharp decrease of
$\rho_{cr}$ in the range $Z/A\sim$ 0.3--0.4. 
The lower curves in each
window correspond to parameters' values satisfying Witten-Bodmer
hypothesis even in the absence of a diquark condensate.  In these
cases, and for $Z/A\sim 0.3$, the critical density is of the order of
$\rho_0$. This opens the possibility to test the deconfinement
transition even in relatively low energy experiments, possible in
future $RNB$ facilities.

The main features  can be easily understood
if one recalls that we are investigating situations in which the minimum
of pure quark matter $EoS$ is at an energy just above or just below
the minimum of the hadronic matter $EoS$. 
The first scenario is the one in which
the absolute minimum of $E/A$,
{\it for a given value of $Z/A$}, corresponds to the quark matter
$EoS$ (this situation corresponds to very small values of the parameter $B$,
e.g. $B^{1/4}=148~MeV$ in Fig.~\ref{fig1}).
In this case,
the deconfinement transition starts at very small densities, even smaller
than nuclear matter saturation density.
The numerical determination of these densities is rather delicate
and we limit ourself to 
indicate with vertical arrows, like in Fig.~\ref{fig1}, the behavior
of the crossing density for such value of $B$. If the value of
$B$ is further reduced, the vertical arrow shifts towards larger values
of $Z/A$ and therefore cannot correspond to a physically acceptable
situation, since it would imply deconfinement 
into two flavor quark matter at low densities, even for almost
symmetric nuclei. 

The second situation is the one in which the minimum of the
quark $EoS$ lies slightly above the hadronic minimum,
as e.g. for $B^{1/4}=150~MeV$ in the top panel of Fig.~\ref{fig1}. 
In this case
the deconfinement transition starts at a density slightly smaller
than the one corresponding to the minimum of the quark $EoS$. 
The crossing density cannot be further reduced,
since at even smaller densities the energy $E/A$
in the quark phase rises dramatically, both in the
$MIT$ bag model and in the CDM, and therefore no mixing of 
hadronic matter with quark matter is possible at those
densities. 

Finally, when the value of $B$ is further increased, 
the isospin dependence of the energy becomes
percentually smaller, the 
dependence of the crossing density on the $Z/A$ fraction reduces
progressively, and a situation similar to the one discussed in
Ref.\cite{MuellerNPA618} is reached.

\begin{figure}
\begin{center}
\includegraphics[angle=+90,scale=0.5]{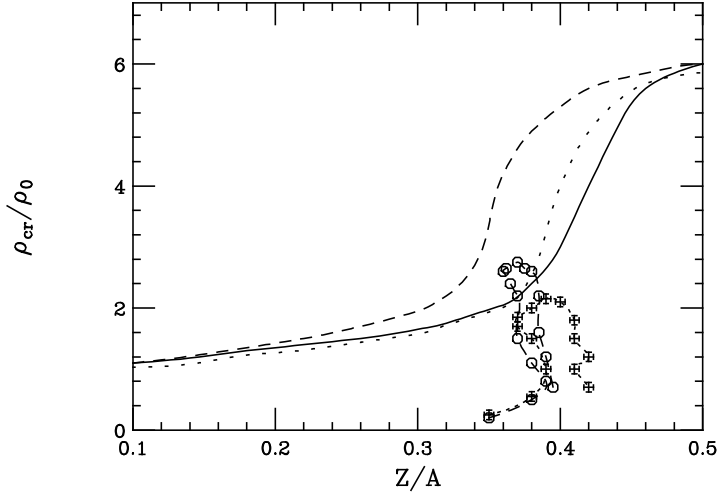}
\caption{\label{rhodelta}
Variation of the transition density with proton fraction for various
hadronic $EoS$ parameterizations. Dotted line: $GM3$ parametrization
\cite{GlendenningPRL18}; dashed line: $NL\rho$ parametrization
\cite{LiuboPRC65}; solid line: $NL\rho\delta$ parametrization
\cite{LiuboPRC65}. For the quark $EoS$, the $MIT$ bag model with
$B^{1/4}$=150 $MeV$ and $\alpha_s$=0 has been used.
The points represent the path followed
in the interaction zone during a semi-central $^{132}$Sn+$^{132}$Sn
collision at $1~AGeV$ (circles) and at $300~AMeV$ (crosses), see text. 
}
\end{center}
\end{figure}

In Fig.\ref{rhodelta} the effect of the exchange of the charged
$\delta$ meson is considered in Non-Linear models \cite{LiuboPRC65}. 
The $\delta$-exchange potential provides an extra isospin dependence
of the $EoS$, and its effect shows up
in a further reduction of the
critical density.

Let us now comment on the physical relevancy of the dramatic
reduction of the crossing density in neutron rich nuclei
at zero temperature.
First, the request that stable neutron-rich nuclei do not dissolve into quarks
puts more stringent bounds on the model parameters than the usual request
based on iron stability.
Second, it is interesting to discuss which could be the signatures of 
the beginning of the formation of mixed phase, at a density of the order of
$\rho_0$, in neutron rich nuclei.
Although we cannot expect to find a 
direct signal of deconfinement in the structure
of these nuclei, we can look for precursor signals.
In particular, we expect that the formation of clusters
containing six or nine quarks will be enhanced due to the reduction
of the crossing deconfinement density. 
This enhancement can in turn be interpreted as a modification of 
single-nucleon properties due the nuclear medium.
The experimental search of effects like the one we are referring to
here has a very long story, which includes the
discovery of the $EMC$ effect \cite{Aubert1983}. This is a non trivial
difference between free-nucleon and nuclear structure functions,
for a review see \cite{ArneodoPR240}. In particular models
invoking the formation of multi-quark clusters have been rather
popular \cite{JaffePRL50,CarlsonPRL51,Barshay2000}.
Our analysis suggests a dependence of the $EMC$ effect
on the isospin, since the probability of forming virtual multi-quark bags
would be enhanced in neutron rich nuclei. This dependence, which has 
a many-body origin, would add to
non-isoscalarity effects which are in any case present in the different
structure functions of neutrons and protons \cite{Eskola1999}.
A full account of in-medium corrections has been attempted in Ref.
\cite{Barone2000}. 
A recent attempt at measuring the isospin dependence
of the $EMC$ effect is reported in Ref.\cite{Oldeman2000}.

The above discussed effect around normal density can take place only for rather special values 
of the model parameters, or for nuclei unrealistically neutron-rich.
In the following, therefore, we will discuss effects associated with larger densities,
reachable during intermediate-energy heavy ion collisions.

\begin{figure}
\begin{center}
\includegraphics[angle=+90,scale=0.33]{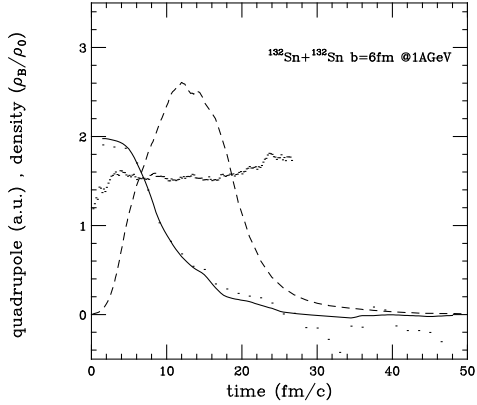}
\caption{\label{sn132cm}
Semicentral $^{132}Sn+^{132}Sn$ collision at $1~AGeV$. 
Time evolution of the quadrupole moment in momentum space
(solid line) and of the density (dashed line). The simulation
examine the after-scattering
thermalization inside a cubic cell $2.5~$fm wide, located in the center
of mass of the system.  
}
\end{center}
\end{figure}

\section{Relativistic Transport Simulations: Exotic Transient States}

A direct way to explore the reduction of the 
deconfinement transition density would be
to test the $EoS$ of asymmetric matter via collisions
of two neutron rich nuclei.
This possibility is based on the ``ab initio'' analysis
of intermediate-energy heavy-ion collisions in a
Relativistic Mean Field approach, as 
discussed in 
Refs.\cite{GrecoPLB562,GaitanosNPA732,GaitanosPLB595,BaranPR410,SantiniNPA756,FeriniNPA762}. 

\subsection*{$^{132}Sn$ Collisions}

We have first performed some simulations of the $^{132}$Sn +
$^{132}$Sn collision (average $Z/A$=0.38) at various energies, for
semicentral impact parameter, $b$=6 $fm$, in order to optimize the neutron
skin effect and get a large asymmetry in the interaction
zone. In order to be fully consistent we have used the same effective
interaction $NL\rho\delta$ \cite{LiuboPRC65} of the $EoS$ leading to
the transition deconfinement density (solid line) of Fig.\ref{rhodelta}.
In the same figure we report the paths in the $(\rho,Z/A)$
plane followed in the c.m. region during the collision, at energies of
$300~AMeV$ (crosses) and $1~AGeV$ (circles). We see that already at
$300~AMeV$ we are reaching the border of the mixed phase, and we are
well inside it at $1~AGeV$.  We have also performed another check of
feasibility.  Since the neutrons in the skin of $^{132}$Sn occupy an
extended region of very low density, it is important to check if
during the collision the nucleons remain near the c.m. when the center
of the system has thermalized. In Fig.\ref{sn132cm} we show that
indeed, when the maximum density is reached ($\rho\sim 2.6\, \rho_0$)
the quadrupole moment of the nucleons momentum distribution has
dropped to $\sim 10\%$ of its initial value, a signal that the system
has indeed thermalized.

The use of an harder hadronic $EoS$ for symmetric matter at high
density would correspond to a even more favorable situation than the
one presented in Fig.\ref{rhodelta}, as discussed before.  Moreover
the use of neutron-richer nuclei would allow to test the $EoS$ at
smaller values of $Z/A$. To this purpose, the most promising nuclei
are the ones near the r-process path, in particular for neutron
numbers near the magic values N=82 or 126.  In these regions, the
proton fraction is as low as 0.32--0.33 and these nuclei could be
studied in future experiments with neutron-rich beams.

\begin{figure}
\begin{center}
\includegraphics[scale=0.37]{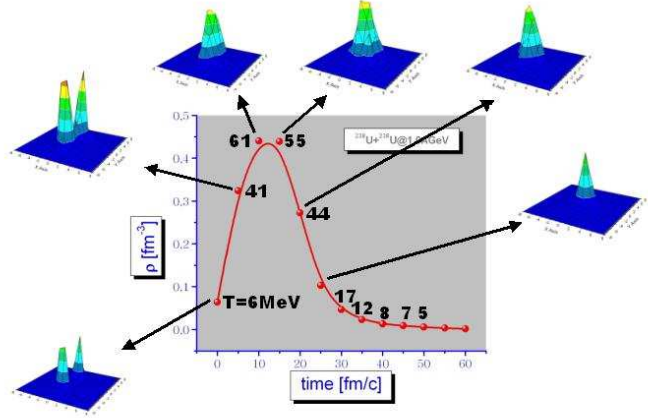}
\caption{\label{figUU1}
Uranium-Uranium $1~AGeV$ semicentral: correlation between density, 
temperature, momentum
thermalization inside a cubic cell 2.5 $fm$ wide, located in the center
of mass of the system.}
\end{center}
\end{figure}

\begin{figure}
\begin{center}
\includegraphics[angle=-90,scale=0.33,trim=0 0 -20 0]{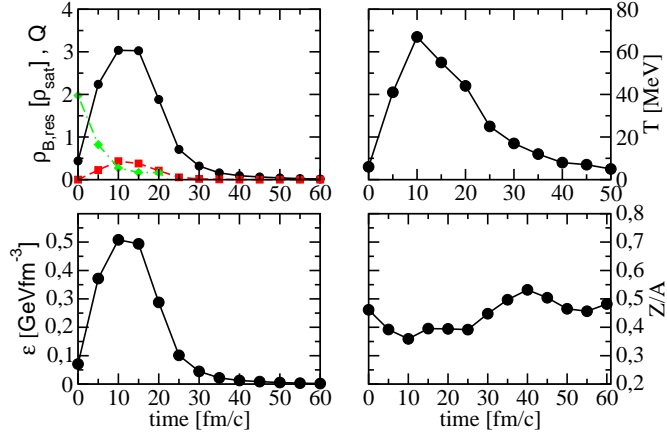}
\caption{\label{figUU2}
Uranium-Uranium $1~AGeV$ semicentral: density, temperature, energy density, 
momentum, isospin inside a cubic cell 2.5 $fm$ wide, located in the center
of mass. Curves in the upper-left panel: {\it black dots} -
baryon density in $\rho_0$ units; {\it grey dots} - quadrupole moment in 
momentum space; {\it squares} - resonance density.  
}
\end{center}
\end{figure}

\subsection*{$^{238}U$ Collisions}

\begin{figure}
\begin{center}
\includegraphics[scale=0.6]{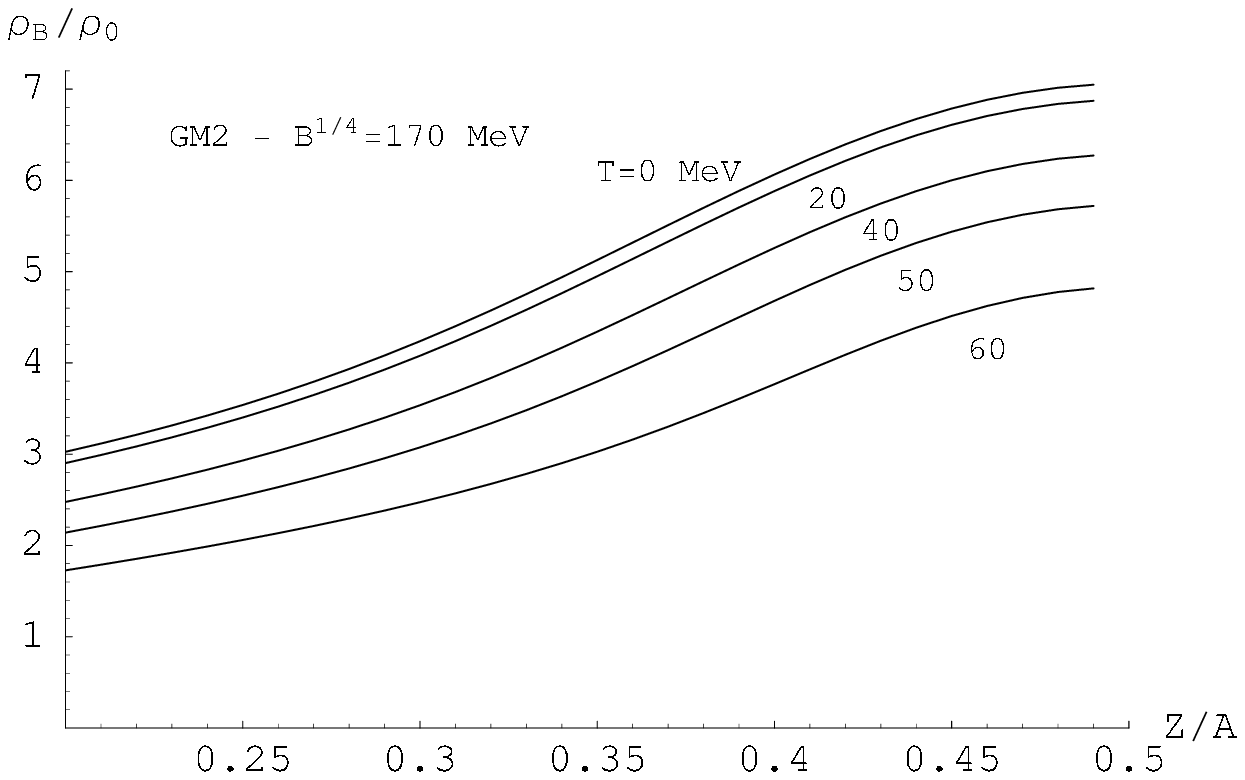}\\
\includegraphics[scale=0.6]{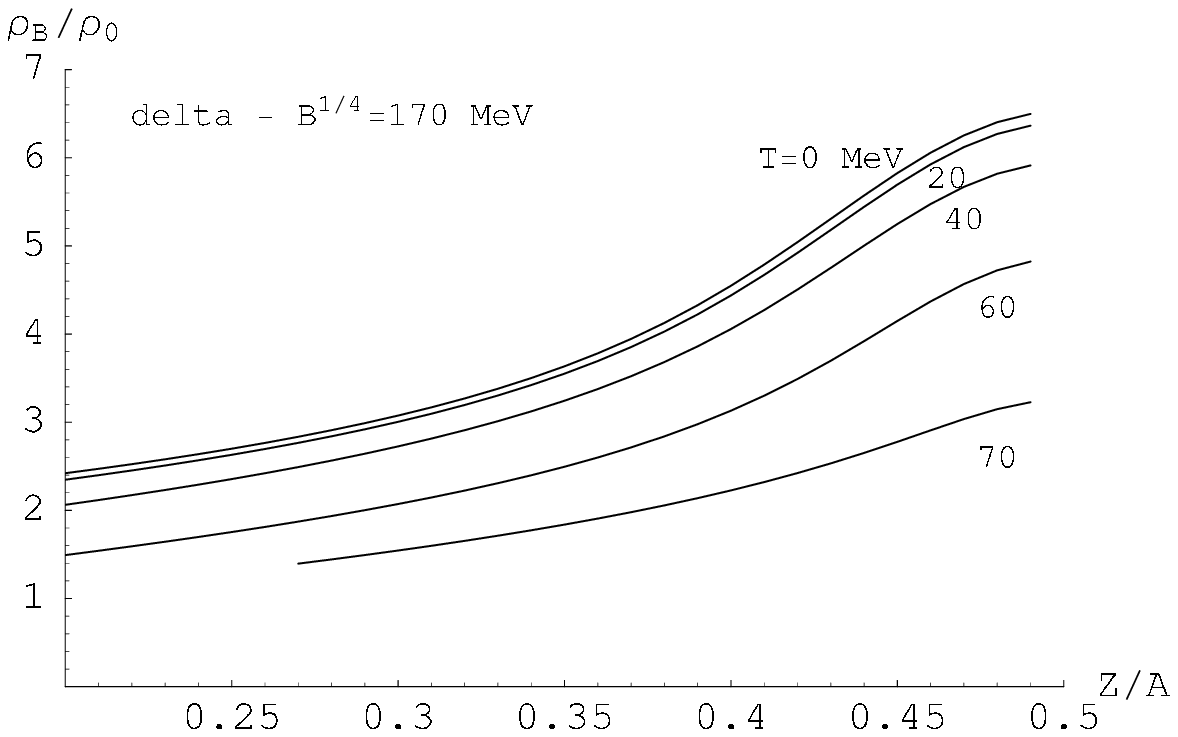}
\end{center}
\caption{\label{temp}
Transition densities separating hadronic matter from mixed
quark-hadron phase at finite temperature. In the upper panel the $GM2$ 
parametrization
has been used for the hadronic $EoS$, in the lower panel, the
$NL\rho\delta$ parametrization.
The $MIT$ bag model with $B^{1/4}=170~MeV$, without
gluon exchange, has been used for the quark $EoS$.
}
\end{figure}

In order to check the possibility of observing some precursor signals
of this new physics even in collisions of stable nuclei at
intermediate energies we have performed some event simulations for the
collision of very heavy, neutron-rich, elements. We have chosen the
reaction $^{238}U+^{238}U$ (average proton fraction $Z/A=0.39$) at
$1~AGeV$ and semicentral impact parameter $b=7~fm$ in order to increase
the neutron excess in the interacting region. We have used a
$NL\rho\delta$ Hadronic Lagrangian in order to optimize the shift of
the transition density at high isospin density. We can compare
directly the conditions of the nuclear matter at local equilibrium
during the reaction evolution with the predictions of the lower panel
of Fig.\ref{temp} obtained with the same interaction.  In order to
evaluate the degree of local equilibration and the corresponding
temperature we have also followed the momentum distribution in a space
cell located in the c.m. of the system, in the same cell we report the
maximum mass density evolution. The results are shown in Fig.~\ref{figUU1}.  
We see that after about $10~fm/c$ a nice local
equilibration is achieved.  We have an unique Fermi distribution and
from a simple fit we can derive a local temperature evaluation. At
this beam energy the maximum density (about three times $\rho_0$)
coincides with the thermalization at estimated maximum temperatures of
$50-60~MeV$, then the system is quickly cooling while expanding.

In Fig.\ref{figUU2} we report the time evolution of all physics
parameters inside the c.m. cell in the interaction region: upper-left
panel, baryon and resonance density and quadrupole deformation in
momentum space; upper-right panel, local temperature evolution;
lower-left panel, energy density; lower-right panel, proton fraction.

We note that a rather exotic nuclear matter is formed in a transient
time of the order of $10~fm/c$, with baryon density around $3\rho_0$,
temperature $50-60~MeV$, energy density $500~MeV~fm^{-3}$ and proton
fraction between $0.35$ and $0.40$, well inside the mixed phase region
of the Fig.~\ref{temp} lower panel, see the next Section.

\begin{figure}
\begin{center}
\includegraphics[scale=0.6]{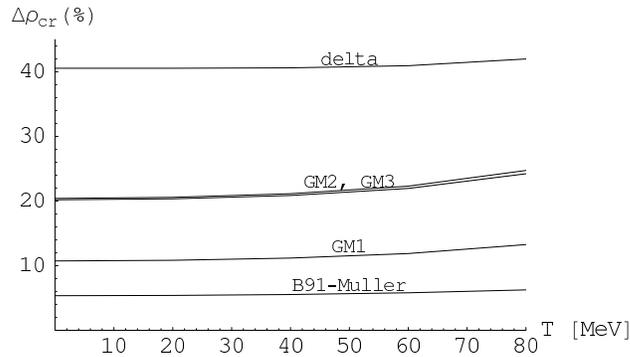}
\end{center}
\caption{\label{deltashift}
Percent variation of the transition density at various temperatures, 
with respect to the symmetric 
case, for a system with initial proton fraction in the hadronic phase
$Z/A=0.35$. Results are shown for 
various choices of the hadronic $EoS$, see text.
The quark $EoS$ is the one used by M\"uller, \cite{MuellerNPA618},
($B^{1/4}=190~MeV$, $\alpha_s=0.35$).}
\end{figure}

\section{Results at Finite Temperature}
\label{finitemp}
As discussed before our aim is finally to suggest possible experiments
on heavy ion collisions at intermediate energies. In a realistic collision
we cannot have a compression of the interacting system without some
heating. This point has been be clearly shown in the Uranium-Uranium
results of the previous Section.

Thus, it is essential to extend the previous results to finite
temperature cases. The procedure is exactly the same as in the $T=0$
case, we only have to use the $T \neq 0$ form of the Fermi
distributions for nucleons and quarks, Eqs.(\ref{eq.7},\ref{eq.8}) and
Eqs.(\ref{nq},\ref{nbarq}) respectively.

The thermal motion reduces the critical baryon density for the
transition to the mixed phase, in agreement with the results obtained
in Ref. \cite{MuellerNPA618}, although this effect is actually more
relevant at higher temperatures, above $T=60~MeV$.  Our results are
shown in Fig.\ref{temp} for $GM2$ (upper panel) and $NL\rho\delta$
(lower panel) hadronic $EoS$. The same quark $EoS$, the $MIT$ bag
model with $B^{1/4}=170~MeV$ and without gluon exchange, has been used
in both cases.

We notice again that the larger symmetry repulsion due to the
inclusion of the $\delta$-meson is increasing the reduction of
$\rho_{tr}$ at any temperature. We show even more explicitly this
effect in Fig.\ref{deltashift} where we report the percent variation
of the transition density respect to the symmetric case, at various
temperatures, for a system having a proton fraction $Z/A=0.35$.

\section{Finite size effects in a Nucleation Mechanism}
\label{finisize}

From the previous simulations it appears that in the few $AGeV$ range,
where the approach presented here is valid, we can just enter the mixed
phase, since the interacting system will always be rather close to the
lower transition density. It is then interesting to discuss the nucleation
mechanism for cluster formation, dominant in the metastable regions of
a first order phase transition.  Thus, in this Section we discuss how
the formation of mixed phase is influenced by finite size effects, due
to a non-vanishing surface tension at the interface between hadronic
($H$) and quark ($Q$) matter.  These effects have been investigated in
the literature both at zero and at finite temperature
\cite{HeiselbergPRL70,OlesenPRD,KutscheraPRC62}.  The crucial quantity
to be studied is the work needed to create a bubble of quark matter of
radius $R$, which reads: 
\bq 
W_{\mathrm
{min}}&=&[F^Q(P_Q)+P_Q V_Q + 4\pi \sigma R^2+
\frac{16\pi^2}{15}(\rho_V^Q-\rho_C^H)^2 R^5]\nonumber\\
&-& [F^H(P_H)+P_H V_H]\,, 
\eq 
where $F$ is the free energy.
The third term in the first bracket describes the energy contribution
of a non-vanishing surface tension $\sigma$, while the fourth term is
related to the Coulomb energy of the drop.  The minimal work can be
rewritten as 
\bq 
W_{\mathrm
{min}}&=&-\frac{4\pi}{3}R^3[(P_Q-P_H) -
\rho_B^Q(\mu_B^Q-\mu_B^H)-\rho_C^Q(\mu_C^Q-\mu_C^H)]\nonumber\\
&+&4\pi \sigma
R^2\frac{16\pi^2}{15} (\rho_C^Q-\rho_C^H)^2 R^5 ,
\label{freeen}
\eq 
where both the baryonic ($B$) and the electric charge ($C$)
appear, since in the here discussed system they are independent
\cite{MuellerPRC52}.  In Fig.\ref{schema} we show the dependence of
$W_{\mathrm {min}}$ on the radius of the drop of quark matter, for two
different values of the baryonic chemical potential of the hadronic
phase, in the case $P_Q=P_H$.

There are three questions which need to be addressed in order to
describe the process of formation of bubbles of quark matter in
heavy-ion scattering experiments: the probability of forming a
critical bubble on a time-scale compatible with the transient times
discussed in Sec.IV, and finally the stability and size of the bubble.

\subsection*{Drop formation rate}
The first question concerns the formation time-scale of drops of quark
matter. The existence of the non-vanishing surface tension produces a
barrier (having a maximum at a radius $R=R_c$) which needs to be
overcome.  At low temperatures the barrier is bypassed via
semi-classical tunneling. At higher temperatures, the ones in which we
are particularly interested in this paper, thermal nucleation is the
relevant process.  The thermal nucleation rate can be estimated as
\cite{OlesenPRD} 
\be 
{\mathcal R}_n=\mu_B^4 \exp (-W_c/T)\label{rate}\, ,
\ee 
where $\mu_B$ is the baryon chemical potential and
$W_c=W_{min}(R_c)$ represents the work needed to form the smallest
bubble capable of growing.  It can be estimated from Eq.(\ref{freeen})
neglecting the differences in the chemical potentials. It corresponds
to the maximum of the free energy of the bubble of quark matter, which
is obtained for a radius $R=R_c$.  Here the prefactor has been related
to the chemical potential, while, when thermal nucleation is
investigated at very large values of the temperature, it is the
temperature which drives the prefactor.  There is a certain ambiguity
on the precise dependence of the prefactor on the chemical potential
and/or the temperature. This ambiguity reflects on the precise
estimate of the nucleation rate, which is anyway dominated by the
exponential factor. We have also explored the possibility of a
prefactor $\sim T\,^4$, obtaining for the critical densities values
larger by roughly $10\%$.  The probability of forming a quark bubble
inside the hot and dense matter produced by high-energy scattering of
two heavy ions is given by 
\be 
{\mathcal P}={\mathcal R}_n\,V_0\,t_0\,\, , 
\ee 
where $t_0$ is the duration of the thermal equilibrium
of the central region, having a volume $V_0$, in which large
temperatures and densities are reached.  The results of the numerical
simulations shown in the previous Section indicate a duration
$t_0\sim$ 10 fm/c and a volume $V_0$ of a few ten fm$^3$.  The
probability ${\mathcal P}$ of forming a critical bubble has not be too
small, if the mixed phase can be produced in a significant fraction of
scattering events.  We will request that ${\mathcal P}$ is of order
unity.

\subsection*{Drop stability}
\begin{figure}[t]
\begin{center}
\includegraphics[scale=0.6]{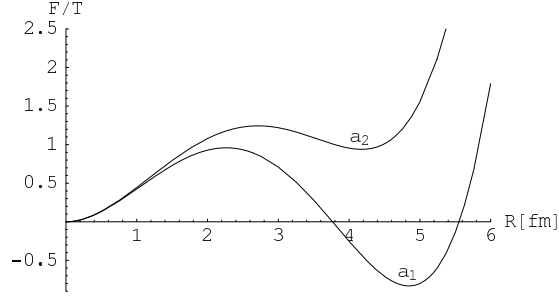}
\end{center}
\caption{
\footnotesize Free energy as a function of the radius of 
the drop. Here Z/A = 0.35, $T$ = 50 $MeV$, B$^{1/4}$ = 160 $MeV$. The upper 
(lower) line
corresponds to a chemical potential $\mu_B$ = 1.075 $GeV$ (1.069 $GeV$).
\label{schema}
}
\end{figure}

Once a bubble having a radius larger than the critical one is formed,
its stability depends on the value of its energy at equilibrium. As it
can be seen from Fig.~\ref{schema}, when the chemical potential is not
large enough a local minimum (a$_2$) does develop at a finite value of
$R$. Therefore, the formed drop is metastable, it decays in a finite
time and no stable mixed phase can exist. Actually, for small
densities and large values of Z/A the local minimum does not exist at
all, due to the large contribution associated with repulsive Coulomb
energy.  For values of the density larger than a critical one, an
absolute minimum exists (a$_1$)
and a stable mixed phase can develop.  The consequences of the
formation of stable droplets have been discussed in the literature in
relation with the value of the critical densities in beta-stable
matter, in connection with the structure of neutron stars
\cite{HeiselbergPRL70}. In that case it is assumed that matter has the
possibility to reach complete chemical and mechanical equilibrium,
because the system is investigated at an asymptotically large
timescale.  Therefore, the structures (drops, ropes and slabs) which
appear in the mixed phase need to be absolutely stable. The outcome of
that analysis is that the critical density, taking into account the
absolute stability of the mixed phase, is larger than the critical
density estimated neglecting finite size effects. In the present
paper, on the other hand, we are not interested in the absolute
stability of the mixed phase, but in the possibility of forming drops
of quark matter that, although metastable, have a life-time long
enough to be observed in high-energy heavy-ion collisions. The crucial
time-scale to which the drops life-time has to be compared is the
duration $t_0$ of the thermal equilibrium of the central region in
which large temperatures and densities are reached during the
scattering process. The numerical results are nevertheless indicating
that the density at which the drop becomes absolutely stable is close
to the density at which the drop is metastable, but with a long enough
life-time. For simplicity, in the following we will therefore consider
only stable drops, what corresponds to a small overestimate of the
critical density.

\subsection*{Drop size}
The final constraint on the minimal density at which the bubble can be
generated comes from the volume of the critical bubble, which
certainly has not to exceed $V_0$.  Actually its size should be
significantly smaller than the size of the central scattering region
and, therefore, the radius of the critical bubble has not to exceed
1.5--2 fm$^3$.

\subsection*{Finite size and finite duration corrections to $\rho_{cr}$ }
The actual value of the minimal density at which mixed phase can be
formed and observed in heavy ion scattering will be determined by the
most restrictive constraint among the ones previously discussed.


\begin{figure}[t]
\begin{center}
\scalebox{0.6}{\includegraphics*[80,510][540,780]{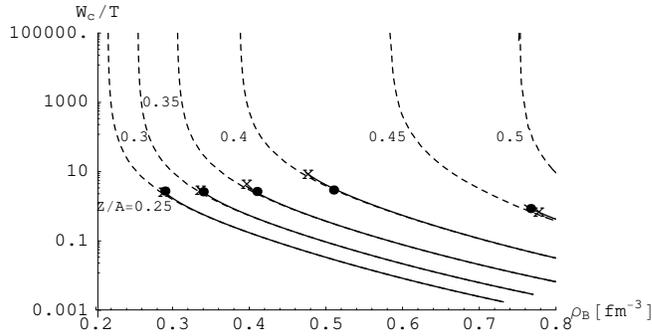}}
\end{center}
\caption{
\footnotesize Work needed to create a drop of quark matter, as a function of 
the density, for various
values of Z/A. 
Here $T$ = 50 $MeV$, B$^{1/4}$ = 160 $MeV$. The solid (dashed) lines include 
(exclude) the Coulomb energy
contribution. See text. \label{zwork160}
}
\end{figure}

\begin{figure}[htb]
\begin{center}
\scalebox{0.5}{\includegraphics*[70,480][550,780]{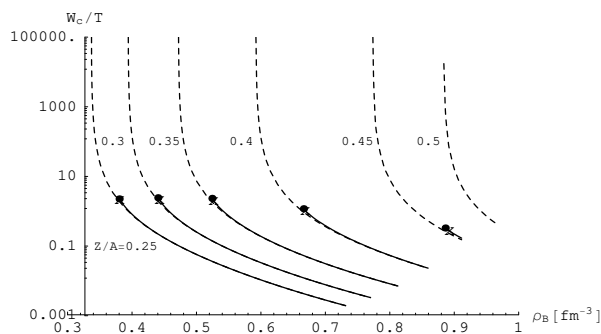}}
\end{center}
\caption{
\footnotesize Same as in Fig.\ref{zwork160}. Here B$^{1/4}$ = 170 $MeV$. 
\label{zwork170}
}
\end{figure}

In Figs.~\ref{zwork160} and \ref{zwork170} we show the work needed to
create a critical size drop at a temperature T=50 $MeV$, as a function
of the density and for various values of Z/A. The value of the surface
tension has been taken to be $\sigma$ = 10 $MeV/fm^2$, as suggested by
microscopic calculations \cite{BergerPRC35} and as adopted in other
works \cite{VoskNPA723,BerezhianiAJ586,DragoPRD71}.  Finally, the
value of the bag pressure is $B^{1/4}=160$ $MeV$ (Fig.\ref{zwork160})
and $B^{1/4}=170$ $MeV$ (Fig.\ref{zwork170}).  As it can be seen, $W_c$
strongly depends on Z/A and also on the density, in a way rather
similar to the one discussed in Ref.\cite{BerezhianiAJ586} where the
nucleation was due to semiclassical tunneling.  The solid lines are
computed using the complete free-energy of Eq.(\ref{freeen}), while
the dashed lines are obtained neglecting the Coulomb term. It is
interesting to notice that, when the Coulomb energy is neglected, the
work $W_c$ diverges at a density $\rho_{Gibbs}$ which corresponds to
the critical (transition) density in the absence of surface tension (i.e. the
density shown in the previous sections), because at $\rho_{Gibbs}$ the
gain in bulk energy vanishes. On the other hand, when the Coulomb term
is taken into account, no local minimum does develop below a density
$\rho_{metastab}$ (corresponding to the end of the solid lines), which
is numerically larger than $\rho_{Gibbs}$.

As previously discussed, we require the nucleation probability
${\mathcal P}$ to be at least of order unity, what fixes the maximal
value of the work to be $W_c/T\sim 10$. In Figs.~\ref{zwork160} and
\ref{zwork170} we also indicate with a black dot the minimal density
$\rho_{size}$, for each value of Z/A, at which the critical radius
$R_c\lesssim 1.8$ $fm$.  Finally, small crosses indicate the density
$\rho_{stab}$ at which the local minimum of the free energy becomes a
global minimum and therefore the drops are absolutely stable.


\begin{figure}[htb]
\begin{center}
\includegraphics[scale=0.7]{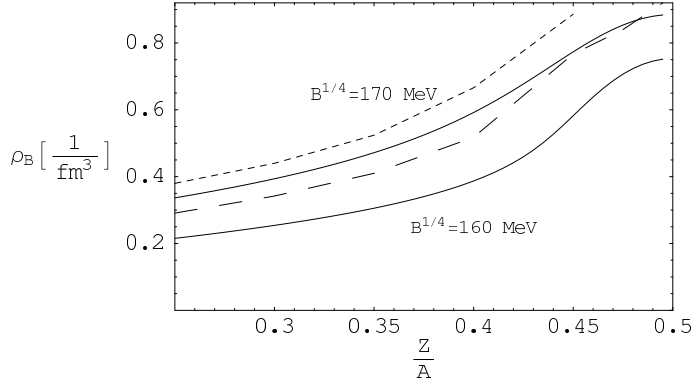}
\end{center}
\caption{
\footnotesize
Critical (transition) densities, as a function of Z/A, at $T=$ 50 $MeV$. 
The solid lines are 
obtained neglecting 
finite-size effects. The long and short-dashed lines take into account these 
effects
and correspond to B$^{1/4}$ = 160 $MeV$ and to B$^{1/4}$ = 170 $MeV$, respectively.
\label{figzsigma}
}
\end{figure}


As it can be seen, while finite size effects increase the value of the
critical density, the various constraints related to the formation and
stability of a drop of quark matter can be satisfied for densities
having a rather similar numerical value.
The less stringent constraint appears to be the one on the nucleation
rate, which is always satisfied once the constraints on the radius and
the stability of the drop are imposed \footnote{It can be interesting 
to notice that, for small values
of $B$, if the proton fraction is less than $0.4$ the main role is played
by the constraint on the maximum radius of the drop. At variance, for
larger values of $Z/A$ the main constraint comes from the existence of
the local minimum since the Coulomb energy becomes larger}.

In Fig.~\ref{figzsigma} we show the impact on the critical (transition) 
densities
of finite size effects.  Clearly the value of the density at which
mixed phase can be observed becomes larger.  The increase in the
value of the transition density due to finite size effects is larger
for smaller values of $B$. The main result of our work, namely a
strong dependence of the transition densities on the value of Z/A, is
confirmed by this analysis.

\subsection*{Statistical fluctuations}

Before closing this Section, it is worth noticing that statistical
fluctuations could help reducing the density at which drops of quark
matter form and can be detected. The reason is that a small bubble can
be energetically favored if it contains quarks whose Z/A ratio is
{\it smaller} than the average value of the surrounding region. This
is again due to the strong Z/A dependence of the free energy, which
favors configurations having a small electric charge. If, for
instance, we consider a bubble having a volume $\sim$ 8 fm$^3$, at the
densities here discussed the bubble contains a baryon number of the
order of 4. It is clear that, although the average value of Z/A can be
significantly larger, random fluctuations can easily produce
configurations in which e.g. 4 neutrons are simultaneously present in
that small volume. These configurations can easily transform into a
bubble of quarks having the same flavor content of the original
hadrons, even if the density of the system is not large enough to
allow deconfinement in the absence of statistical fluctuations. It is
not easy to numerically quantify the relevance of these fluctuations,
but we can conclude that, at densities intermediate between the ones
corresponding to the neglect of finite size effects and the ones in
which those effects are taken into account, precursory signals of
formation of drops of quark matter should appear, if statistical
fluctuations are taken into account. Moreover, since statistical
fluctuations favor the formation of bubbles having a smaller Z/A,
neutron emission from the central collision area should be suppressed,
what could give origin to specific signatures of the mechanism
described in this paper. This corresponds to a {\it neutron trapping}
effect, supported also by the difference in the symmetry energy in the
two phases, which will be discussed in the next section.

\section{Deconfinement Precursors}
\label{precursors}

If the transition occurs, we can expect in general a softening of the
nuclear $EoS$, but this can be accounted for even in a pure hadronic
picture, i.e. inserting some density dependence in the $RMF$ effective
meson couplings.  Here we would like to suggest a possible effect
which more strictly characterizes the transition model discussed above:
a {\it neutron trapping} (or ``neutron distillation'') to the quark 
deconfined clusters.  The
neutron trapping effect corresponds to the formation of a drop of
quarks obtained by deconfining an hadronic drop made mainly by
neutrons.  The physics behind this isospin migration from the hadron
phase to the quark phase is related to the different importance of the
symmetry energy, namely, while in the hadron phase we have a large
potential repulsion (in particular in the $NL\rho\delta$ case) in the
quark phase we only have the much smaller kinetic contribution.  In the
high density region this effect could be rather relevant: while in a
pure hadronic phase neutrons are quickly emitted, when the mixed phase
starts forming neutrons are kept in the interacting system up to the
subsequent hadronization in the expansion stage.

\begin{figure}
\begin{center}
\scalebox{0.6}{\includegraphics*[50,510][510,780]{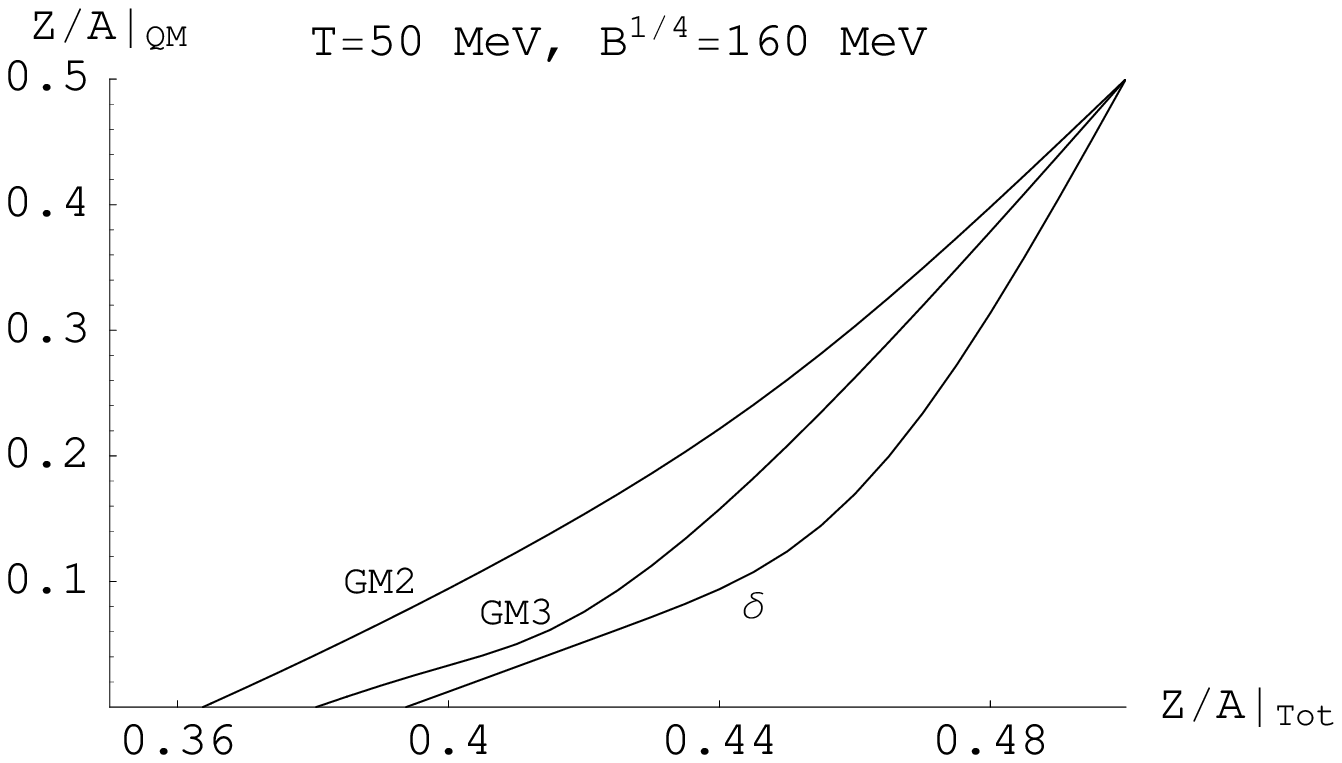}}\\
\scalebox{0.5}{\includegraphics*[70,470][540,780]{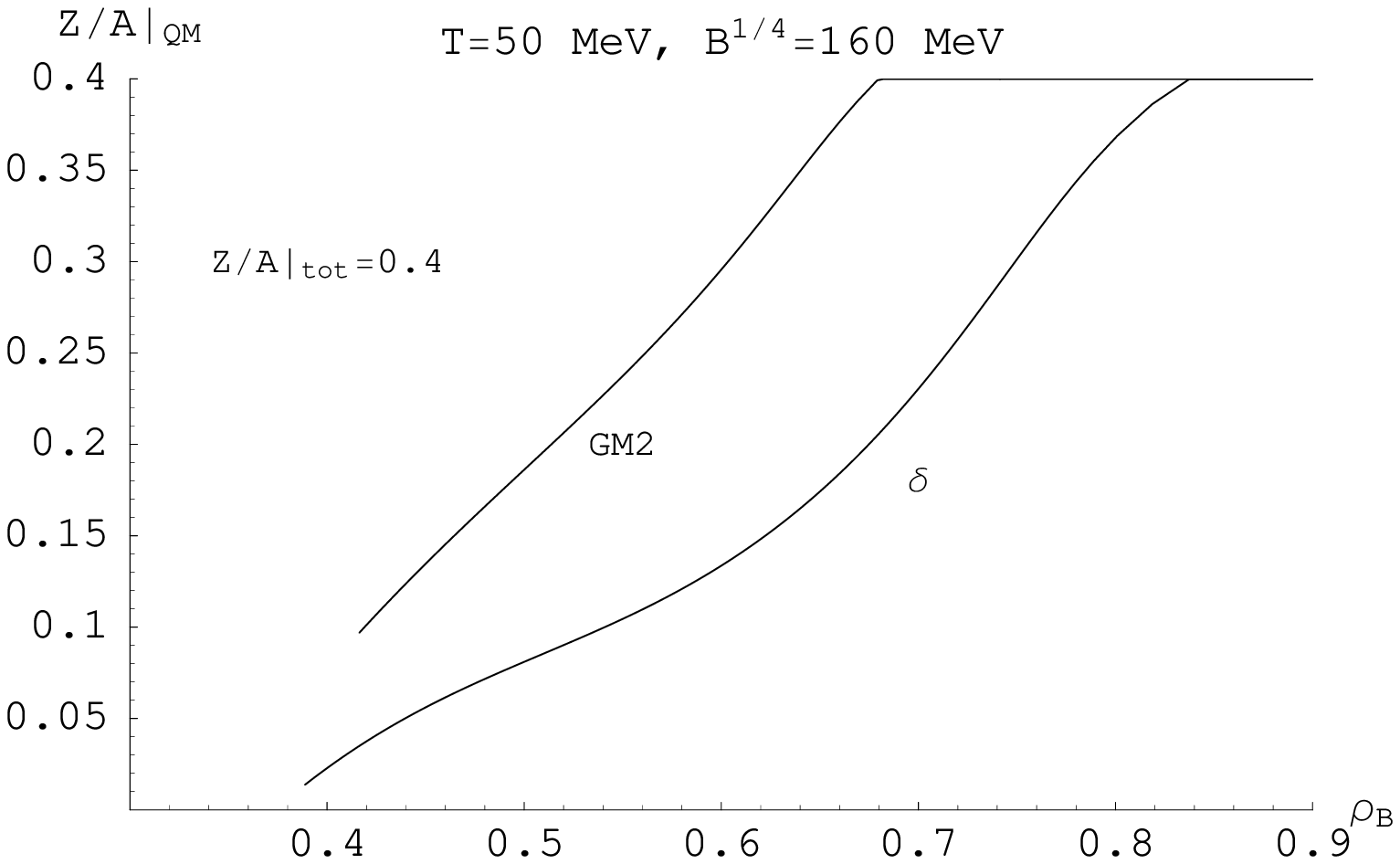}}
\end{center}
\caption{\label{asyquark}
Isospin migration to the quark clusters. The calculations are
performed at a temperature $T=50~MeV$ for various hadron effective
interactions, as indicated. The $MIT$ bag model without gluon exchange
has been used for the quark $EoS$, with a parameter $B^{1/4}=160~MeV$.
Upper panel: Proton fraction of the quark phase {\it at the transition
density}.  Lower panel: Evolution of the proton fraction of the quark
clusters in the mixed phase. Here the initial proton fraction of the
hadronic matter is $Z/A=0.4$.}
\end{figure}

Let us introduce the {\it proton fraction} of a quark drop in the
mixed-phase as the proton content of the hadron drop from which the
quark drop has formed. Clearly, the minimum value of the {\it proton
fraction} is zero, and it corresponds to a drop of quarks produced by
the melting of only neutrons.  The very low {\it proton fraction} of
the quark clusters formed just at the transition density is reported
in the upper panel of Fig.\ref{asyquark} for various types of hadronic
interaction.  In the lower panel of the same figure we report the
dependence on the baryon density of the {\it proton fraction} of the
quark clusters in the mixed phase, for a fixed initial isospin content
of hadronic matter. Of course, at the upper limit of the mixed phase
the initial proton fraction has to be recovered.

Similar results are shown in Tab.~3 where we present
the variation of the proton fraction in the quark clusters inside the
mixed phase for various initial asymmetries of the hadronic
matter. The quantity $\chi$ represent the fraction of quark matter. We
see a decrease of the transition density with the increase of the initial
asymmetry, from $\rho_{tr}=0.58~fm^{-3}$ for $Z/A=0.45$ to
$\rho_{tr}=0.3~fm^{-3}$ for $Z/A=0.35$.  The calculations are performed
at $T=50~MeV$ with the $NL\rho\delta$ hadron interaction and with 
$B^{1/4}=160~MeV$.

\begin{table}[htb]
\begin{center}
\centerline{{\bf Table 3.}~Proton Fraction in the Quark Phase.}

\begin{tabular}{c c c c c c} 
\multicolumn{3}{c}{$Z/A=0.35$}
           &\multicolumn{3}{c}{$Z/A=0.38$} \\
$\rho(\fmr^{-3})$  &$\chi$   &$Z/A|_q$ &$\rho(\fmr^{-3})$  &$\chi$   
&$Z/A|_q$ \\ \hline
 0.3  & 0.0    & -0.08    & 0.35      & 0.0   & -0.02 \\ \hline
 0.38 & 0.1    &  0.0     & 0.37      & 0.03  &  0.0  \\ \hline
 0.54 & 0.27   &  0.1     & 0.54      & 0.18  &  0.1  \\ \hline
 0.68 & 0.42   &  0.2     & 0.68      & 0.31  &  0.2  \\ \hline
 0.77 & 0.70   &  0.3     & 0.76      & 0.53  &  0.3  \\ \hline
\hline
\multicolumn{3}{c}{$Z/A=0.42$}
           &\multicolumn{3}{c}{$Z/A=0.45$} \\ 
$\rho(\fmr^{-3})$  &$\chi$   &$Z/A|_q$ &$\rho(\fmr^{-3})$  &$\chi$   
&$Z/A|_q$ \\ \hline
 0.44  & 0.0    & 0.05    & 0.58      & 0.0   & 0.12 \\ \hline
 0.54 & 0.06    &  0.1     & 0.67      & 0.06  &  0.2  \\ \hline
 0.67 & 0.16   &  0.2     & 0.73      & 0.15  &  0.3  \\ \hline
 0.74 & 0.31   &  0.3     & 0.78      & 0.40  &  0.4  \\ \hline
\end{tabular}

\end{center}
\end{table}

In conclusion we expect such neutron distillation effect to be
particularly efficient if the system is just entering the mixed phase
region. Observables related to such neutron ``trapping'' could be an
inversion in the trend of the formation of neutron rich fragments
and/or of the $\pi^-/\pi^+$, $K^0/K^+$ yield ratios for reaction
products coming from high density regions, i.e. with large transverse
momenta.  In general we would expect a modification of the rapidity
distribution of the emitted ``isospin'', with an enhancement at
mid-rapidity joint to large event by event fluctuations.
A more detailed analysis of these observables is clearly needed
and it will be performed in future investigations.

\section{Perspectives}
\label{perspect}
In our work we have investigated the possible formation of a mixed phase
of hadrons and quarks during intermediate-energy collisions between
neutron-rich heavy ions.  In particular we have examined the mechanism
of production of a drop of quark matter in the central region, where
densities of a few time $\rho_0$ and temperatures of a few ten $MeV$ can
be reached.  This is the ideal scenario for testing the formation of a
mixed quark-hadron phase, a state somehow similar to the one that
could be present in the center of compact stars. An important
difference with the compact star scenario is the very low value of
strangeness present in the scattering central region. In fact, since
weak-decays cannot take place during the short life-time of the high
density system, the only possibility of producing strangeness is
through associated production but, in the scenario we are discussing,
this process has been shown to be very inefficient
\cite{FeriniNPA762,FuchsPPNP56}.  In the present work we have 
therefore completely
neglected the strangeness production.

The $EoS$ tested in the experiments here discussed shares a rather
strict relation with the one to be used {\it during} the pre-supernova
collapse \cite{Arnett96}.  There, $Z/A\sim 0.34$ \footnote{Due to
neutrino trapping, the total electron lepton fraction is larger,
$Y_{l_e}\sim 0.38$.} and the temperature reached just before the
bounce is of the order of a few ten $MeV$.  Moreover, due to neutrino
trapping, weak processes are suppressed and typically hyperon
production is delayed till the proto-neutron star stars deleptonizing.
A difference with the heavy-ion scattering scenario is that during the
gravitational collapse matter follows an isoentropic $EoS$, while in our
paper we have discussed isothermal $EoS$s. On the other hand, once a
quasi-equilibrium configuration is reached the path previously
followed by the thermodynamical variables is irrelevant.

It has been suggested several times that, if a mixed phase forms
during the gravitational collapse, this would help Supernovae to
explode, due to the softening of the $EoS$
\cite{MigdalPLB83,TakaharaPLB156,GentileApJ414,CoopersteinNPA556,DragoJPG25}.
In our paper we have shown that if indeed mixed phase forms during the
gravitational collapse, then it is likely that signatures of a mixed
phase of hadrons and quarks appear in intermediate-energy scattering
between neutron-rich nuclei.  The total absence of such signatures
would presumably rule out the possibility of deconfinement during the
gravitational collapse, although it would definitively not exclude
deconfinement at a later stage, when $Z/A$ drops to $\sim 0.1$ and the
central density of the neutron star reaches values of several times
$\rho_0$. 
Notice anyway that the evidence of the beginning of a phase transition
in relativistic heavy ion collisions would have important implications
for supernova explosions only if the mixed phase starts appearing
at a density of the order of $2 \rho_0$, or smaller.

In conclusion, our analysis supports the possibility of observing
precursor signals of the phase transition to deconfined quark matter
at high baryon density in the collision, central or semi-central, of
neutron-rich heavy ions in the energy range of a few $GeV$ per
nucleon. A possible signature could be revealed through an earlier
softening of the hadronic $EoS$ for large isospin asymmetries, and it
would be observed e.g. in the behavior of the collective flows.  We also
suggest to look at observables particularly sensitive to the
expected different isospin content of the two phases, which leads to a
neutron trapping in the quark clusters.
The isospin structure of hadrons produced at high transverse momentum
should be a good indicator of the effect.


\begin{thebibliography}{00}

\bibitem{MuellerNPA618} H.Mueller, Nucl.Phys. {\bf A618}, 349 (1997).

\bibitem{GlendenningPRL18} N.K.Glendenning, S.A.Moszkowski, Phys.Rev.Lett. 
 {\bf 67}, 2414 (1991).

\bibitem{DanielNPA673} P.Danielewicz, Nucl.Phys. {\bf A673}, 375 (2000).

\bibitem{DanielSCI298} P.Danielewicz, R.Lacey and W.G.Lynch, 
 Science {\bf 298},  1592 (2002).

\bibitem{BrockmannPRC42} R.Brockmann, R.Machleidt,
            Phys.Rev. {\bf C42}, 1965 (1990).

\bibitem{GrossNPA648} T.Gross-Boelting, C.Fuchs, A.Faessler,
            Nucl.Phys. {\bf A648}, 105 (1999).

\bibitem{HaarPR149} B.ter Haar and R.Malfliet,
            Phys.Rep. {\bf 149}, 207 (1987).

\bibitem{GaitanosEPJA12} T.Gaitanos, C.Fuchs, H.H.Wolter, and A.Faessler,
            Eur.Phys.J. {\bf A12}, 421 (2001).

\bibitem{GrecoPLB562} V.Greco et al., Phys.Lett. {\bf B562}, 215 (2003).

\bibitem{GaitanosNPA732} T.Gaitanos et al., Nucl.Phys. A{\bf 732}, 24 (2004).

\bibitem{GaitanosPLB595} T.Gaitanos, M.Colonna, M.Di Toro and H.H.Wolter,\\
 Phys.Lett. {\bf B595 }, 209 (2004).

\bibitem{LiuboPRC65} B.Liu, V.Greco, V.Baran, M.Colonna, and M.Di Toro,\\
            Phys.Rev. {\bf C65}, 045201 (2002).

\bibitem{MitbagPRD9} A.Chodos et al., Phys.Rev. {\bf D9}, 3471 (1974).

\bibitem{FarhiPRD30} E.Farhi, R.L.Jaffe, Phys.Rev. {\bf D30}, 2379 (1984).


\bibitem{Birse1990} M.C.Birse, Prog.Part.Nucl.Phys. {\bf 25}, 1 (1990).

\bibitem{Pirner1992} H.J. Pirner, Prog.Part.Nucl.Phys. {\bf 29}, 33 (1992).

\bibitem{Drago1995} A.Drago, M.Fiolhais, U.Tambini, Nucl.Phys. {\bf A588},
 801 (1995).


\bibitem{AlcockAJ310} C.Alcock, E.Fahri, A.Olinto, Astrophys.J. {\bf 310}, 
 261 (1986).

\bibitem{HaenselAA160} P.Haensel, J.L.Zdunik, R.Schaeffer, Astron.Astrophys.
 {\bf 160}, 121 (1986).

\bibitem{DragoPLB511} A.Drago, A.Lavagno, Phys.Lett. {\bf B511}, 229 (2001).

\bibitem{WittenPRD30} E.Witten, Phys.Rev. {\bf D30}, 272 (1984).

\bibitem{BodmerPRD4} A.R.Bodmer, Phys.Rev. {\bf D4}, 1601 (1971).              
 
\bibitem{AlfordPRD67} M.Alford, S.Reddy, Phys.Rev. {\bf D67}, 074024 (2003).

\bibitem{DragoPRD69} A.Drago, A.Lavagno, G.Pagliara, Phys.Rev. {\bf D69},
 057505 (2004).
                                                            .
\bibitem{LiPRL83} X.D.Li et al., Phys.Rev.Lett. {\bf 83}, 3776 (1998).

\bibitem{DeyPLB438} M.Dey et al., Phys.Lett. {\bf B438}, 123 (1998).

\bibitem{PonsAJ564}  A.Pons et al., Astrophys.J. {\bf 564}, 981 (2002).

\bibitem{DrakeAJ572} J.J.Drake et al., Astrophys.J. {\bf 572}, 996 (2002).

\bibitem{Landaustat} L.D.Landau, L.Lifshitz, {\it Statistical Physics},
 Pergamon Press, Oxford 1969.

\bibitem{GlendenningPRD46} N.K.Glendenning, Phys.Rev. {\bf D46}, 1274 (1992).

\bibitem{KubisPLB399} S.Kubis, M.Kutschera,
             Phys.Lett. {\bf B399}, 191 (1997).

\bibitem{GrecoPRC67}V.Greco, M.Colonna, M.Di Toro, F.Matera,
            Phys.Rev. {\bf C67}, 015203 (2003).


\bibitem{FeriniNPA762} G.Ferini, M.Colonna, T.Gaitanos, M.Di Toro, Nucl.Phys.
 {\bf A762}, 147 (2005).

\bibitem{AmorePRD65} P.Amore, M.C.Birse, J.A.McGovern, N.R.Walet, Phys.Rev. 
{\bf D65}, 074005 (2002).

\bibitem{glendenningbook} N.K.Glendenning, {\it Compact Stars}, Springer-Verlag, New York 1997. 

\bibitem{neuber} T.Neuber, M.Fiolhais, K.Goeke, J.N.Urbano,
  Nucl.Phys. {\bf A560}, 909 (1993).

\bibitem{Aubert1983} J.J.Aubert et al., Phys.Lett. {\bf B123}, 275 (1983).

\bibitem{ArneodoPR240} M.Arneodo, Phys.Rep. {\bf 240}, 301 (1994).

\bibitem{JaffePRL50}  R.L.Jaffe, Phys.Rev.Lett. {\bf 50}, 228 (1983).

\bibitem{CarlsonPRL51} C.E.Carlson, T.J.Havens, Phys.Rev.Lett. {\bf 51}, 
 261 (1983).

\bibitem{Barshay2000} S.Barshay, G.Kreyerhoff, Phys.Lett. {\bf B487}, 
 341 (2000).

\bibitem{Eskola1999} K.J.Eskola, V.J.Kolhinen, C.A.Salgado, Eur.Phys.J. 
 {\bf C9}, 61 (1999).

\bibitem{Barone2000} V.Barone, C.Pascaud, F.Zomer,  Eur.Phys.J. 
 {\bf C12}, 243 (2000)\\
Proc. $DIS2000$, Liverpool, World Sci. 2000, p.162.

\bibitem{Oldeman2000} A.Kayis-Topaksu {\it et al.} (CHORUS Collab.), Eur.Phys.J. {\bf C30}, 159 (2003).

\bibitem{BaranPR410} V.Baran, M.Colonna, V.Greco, M.Di Toro, 
 Phys.Rep. {\bf 410}, 335-466 (2005).

\bibitem{SantiniNPA756} E.Santini, T.Gaitanos, M.Colonna, M.Di Toro,
 Nucl.Phys. {\bf A756}, 468 (2005).

\bibitem{HeiselbergPRL70} H.Heiselberg, C.J.Petick, E.F.Staubo,
 Phys.Rev.Lett. {\bf 70}, 1355 (1993).

\bibitem{OlesenPRD} M.L.Olesen and J.Masden, Phys.Rev. {\bf D47},
 2313 (1993) and  Phys.Rev. {\bf D49}, 2698 (1994)

\bibitem{KutscheraPRC62} M.Kutschera and J.Niemiec, Phys.Rev. {\bf C62}, 
 025802 (2000).

\bibitem{MuellerPRC52} H.Mueller and B.D.Serot, Phys.Rev. {\bf C52},
 2072 (1995).

\bibitem{BergerPRC35} M.S.Berger and R.L.Jaffe, Phys.Rev. {\bf C35}, 
 213 (1987).

\bibitem{VoskNPA723} D.N.Voskresensky, M.Yasuhira, T.Tatsumi, Nucl.Phys.
 {\bf A723}, 291 (2003).

\bibitem{BerezhianiAJ586} Z.Berezhiani, I.Bombaci, A.Drago, F.Frontera,
 A.Lavagno, Astrophys.J. {\bf 586}, 1250 (2003).

\bibitem{DragoPRD71} A.Drago, A.Lavagno, G.Pagliara, Phys.Rev. {\bf D71},
 103004 (2005).

\bibitem{FuchsPPNP56} C.Fuchs, Prog.Part.Nucl.Phys. {\bf 56}, 1-103 (2006). 

\bibitem{Arnett96} D.Arnett, {\it Supernovae and Nucleosynthesis}, Princeton Univ. Press,
Princeton 1996.

\bibitem{MigdalPLB83} A.B.Migdal, A.I.Cherenoutsan, I.N.Mishustin, Phys.Lett. {\bf B83}, 158 (1979).

\bibitem{TakaharaPLB156} M.Takahara and K.Sato, Phys.Lett. {\bf B156}, 17 (1985).

\bibitem{GentileApJ414} N.A.Gentile {\it et al.}, Astrophys.J. {\bf 414}, 701 (1993).

\bibitem{CoopersteinNPA556} J.Cooperstein, Nucl.Phys. {\bf A556}, 237 (1993).

\bibitem{DragoJPG25} A.Drago and U.Tambini, J.Phys. {\bf G25}, 971 (1999).


\end{thebibliography}


\end{document}